\documentclass[aps,prx,reprint,superscriptaddress]{revtex4-2}

\usepackage{graphicx}
\usepackage{xcolor}
\usepackage{amsmath}
\usepackage{amssymb}
\usepackage[percent]{overpic}

\begin{document}

\def\({\left(}
\def\){\right)}
\def\av#1{\left\langle #1\right\rangle}
\def\ad{a^\dagger}
\def\a2d{a^{\dagger 2}}
\def\b2d{b^{\dagger 2}}
\def\d#1{#1^\dagger}
\def\bydef{\stackrel{\wedge}=}
\def\eq#1{Eq.~(\ref{eq:#1})}
\def\Eq#1{Equation~(\ref{eq:#1})}
\def\eqs#1#2{Eqs.~(\ref{eq:#1}) \& (\ref{eq:#2})}
\def\eqlist#1#2{Eqs.~(\ref{eq:#1}-\ref{eq:#2})}
\def\Eqs#1#2{Equations~(\ref{eq:#1}) \& (\ref{eq:#2})}
\def\Eqlist#1#2{Equations~(\ref{eq:#1}-\ref{eq:#2})}
\def\sec#1{Section~\ref{sec:#1}}
\def\secs#1#2{Sections~\ref{sec:#1} \& \ref{sec:#2}}
\def\fig#1{Fig.\ref{fig:#1}}
\def\figs#1#2{Figs.\ref{fig:#1} \& \ref{fig:#2}}
\def\Fig#1{Figure~\ref{fig:#1}}
\def\Figs#1#2{Figures~\ref{fig:#1} \& \ref{fig:#2}}
\newcommand\tab[1]{Table~\ref{tab:#1}}
\newcommand\bra[1]{\left\langle\,#1\,\right|} %\mid doesn't work with \left, \right
\newcommand\ket[1]{\left|\,#1\,\right\rangle}
\newcommand\scalprod[2]{\left\langle\,#1\,\right|\left.#2\,\right\rangle}
\newcommand\om[1]{\omega_{\scriptscriptstyle #1}}
\newcommand\tr[1]{{\rm Tr}[\mathbf #1]}
\newcommand{\op}[1]{\textcolor{red}{[OP: #1]}}
 % Olivier's comments

%\graphicspath{{Figures/}}

\title{Hypercubic cluster states in the phase modulated quantum optical frequency comb}

\author{Xuan Zhu}
\author{Chun-Hung Chang}
\author{Carlos Gonz\'alez-Arciniegas}
\affiliation{Department of Physics, University of Virginia, 382 McCormick Road, Charlottesville, VA 22903, USA}
\author{Avi Pe'er}
\affiliation{Department of Physics and BINA Center of Nano-technology, Bar-Ilan University, 52900 Ramat-Gan, Israel}
\author{Jacob Higgins}
\affiliation{Department of Physics, University of Virginia, 382 McCormick Road, Charlottesville, VA 22903, USA}

\author{Olivier Pfister}
\affiliation{Department of Physics, University of Virginia, 382 McCormick Road, Charlottesville, VA 22903, USA}
\email[olivier.pfister@gmail.com]{olivier.pfister@gmail.com}

\date{\today}

\begin{abstract}
We propose and fully analyze the simplest technique to date to generate light-based universal quantum computing resources, namely 2D, 3D and, in general, $n$-hypercubic cluster states. The technique uses two standard optical  components: first, a single optical parametric oscillator pumped below threshold by a monochromatic field, which generates Einstein-Podolsky-Rosen entangled states, a.k.a. two-mode-squeezed states, over the quantum optical frequency comb; second, phase modulation at frequencies multiple of the comb spacing (via RF or optical means). The unprecedented compactness of this technique paves the way to implementing quantum computing on chip using quantum nanophotonics.
\end{abstract}

\pacs{}

\maketitle

%%%%%%%%%%%%%%%%%%%%%%%%%%  body  %%%%%%%%%%%%%%%%%%%%%%%%%%
{\em Introduction.} For quantum information technology to become a reality, quantum engineering must come of age. A promising  approach is quantum photonics, marrying fundamental quantum optics with integrated photonics.
%Quantum combs in time and frequency
On the fundamental side, quantum optics provides a scalable platform for continuous-variable (CV) universal quantum computing (QC), based on qumodes (e.g.\ quantum optical fields) rather than qubits ~\cite{Pfister2019,Pfister2004,Menicucci2006,Menicucci2008,Flammia2009}, as has been based demonstrated with temporal or spectral phase-locked quantum optical combs, emitted by optical parametric oscillators (OPOs). The interference of shifted, two-mode-squeezed quantum optical combs has been shown to yield cluster states~\cite{Briegel2001,Zhang2006}, which are universal quantum computing resources~\cite{Raussendorf2001,Menicucci2006}, in the spectral domain~\cite{Pysher2011,Chen2014,Roslund2014} and in the temporal domain~\cite{Yokoyama2013,Yoshikawa2016,Asavanant2019,Larsen2019}, with thousands to millions of entangled qumodes. Other schemes, based on the spatial degrees of freedom have been been accomplished~\cite{Armstrong2012} or proposed~\cite{Pooser2014,Yang2020}. It is important to  note that CVQC can be made fault tolerant at reachable squeezing levels~\cite{Menicucci2014ft,Fukui2018}. 

Most of the aforementioned work in the temporal and spectral domains relied on interfering two to four straddling squeezed quantum combs, as originally proposed in Refs.~\citenum{Menicucci2011a} and also  \citenum{Wang2014a}. In this paper, we show that a single comb is in fact sufficient to generate $n$-hypercubic cluster states of arbitrary dimension $n$. Such states are universal resources for quantum computing for $n=2$~\cite{Raussendorf2001}, For $n=3$, they can also allow quantum error correction topological encoding~\cite{Raussendorf2006}.

Our implementation uses the spectral domain, i.e., the quantum optical frequency comb (QOFC) emitted by a single OPO that is phase modulated by a sparse discrete spectrum (either inside or outside of the OPO cavity).\footnote{As we'll show below, the phase modulator can be integrated in the OPO itself.}  Remarkably, the dimension of the cluster graph is determined by the number of modulation frequencies and its size is determined by their spacing. 

This discovery was made possible by a general analysis of the generated Gaussian graph state, factoring in concrete experimental parameters such as finite squeezing, pump amplitude, and modulation depth~\cite{Gonzalez2020}. Our results account for all graph errors and allow us to drastically simplify experimental configurations, paving the way to compact  realizations of large-scale cluster entanglement using a single OPO on chip~\cite{Furst2011,Dutt2015,Lenzini2018,Mondain2019,Vaidya2020}.

The paper is organized as follows. In \sec{theory}, we recall the quantum description of phase modulation and outline the derivation of the quantum state obtained for a two-mode-squeezed input. In \sec{graphs}, we present our theoretical results and show that a single phase-modulated OPO can generate precisely defined, $n$-hypercubic cluster states. In \sec{error} we thoroughly analyze graph errors and the full effects of finite squeezing. We then conclude.

\section{Phase modulation of Gaussian states}\label{sec:theory}

\subsection{Quantum optical phase modulation}\label{sec:QOPM}

\begin{figure}[t]
\includegraphics[width=\columnwidth]{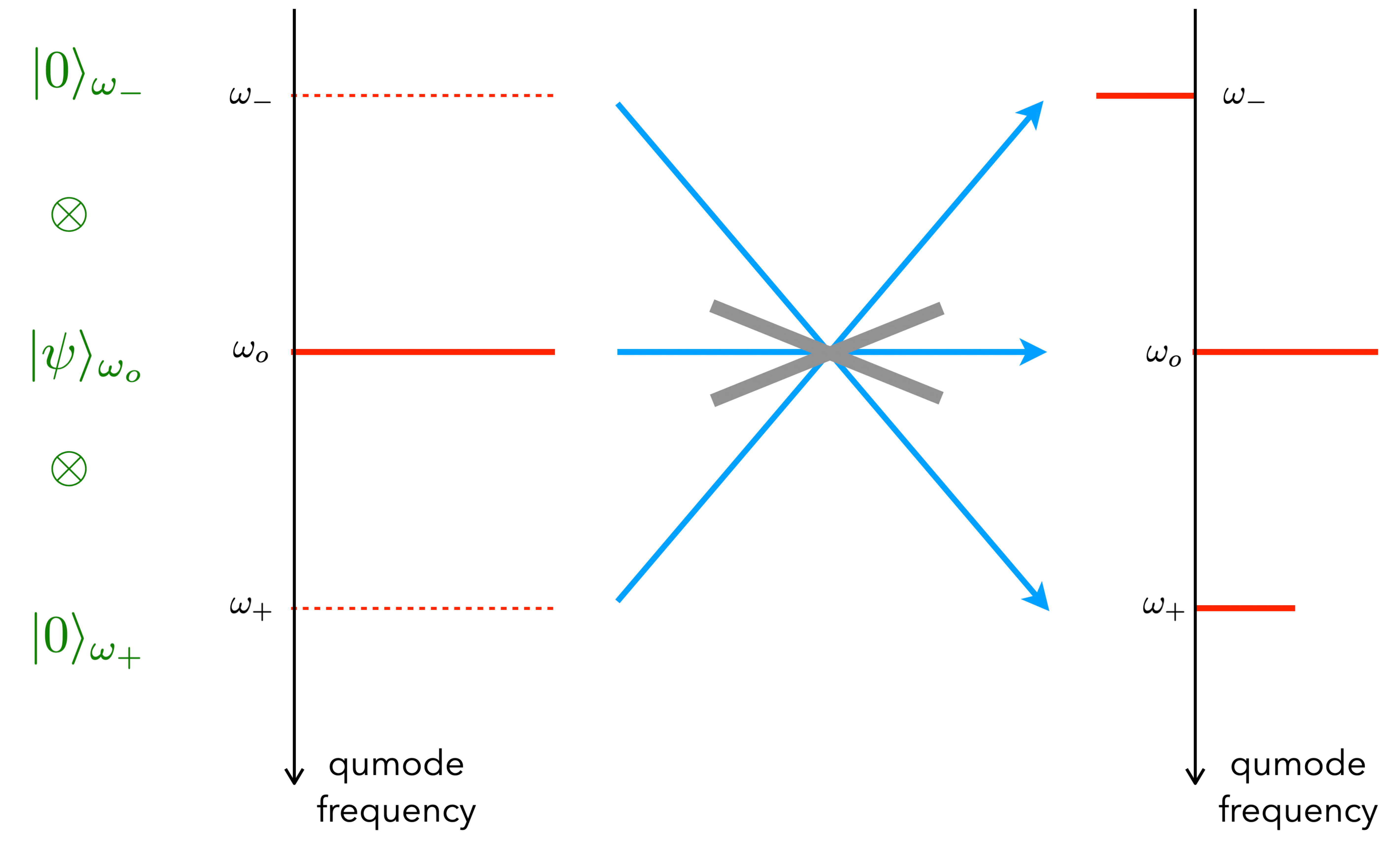}
\vglue-0.1in
\caption{Quantum model of phase modulation at frequency $\Omega$ by a 3$\times$3 frequency-domain beamsplitter, \eq{H1}~\cite{Capmany2010}, symbolized by the unphysical gray element that features two  frequency-domain ``reflective,'' and one ``transmissive,'' possibilities for each input beam (blue lines) at $\om{\pm}$=$\om o$$\pm$$\Omega$. For simplicity, we didn't draw modulation sidebands of order greater than 1. An arbitrary input state is given in green, containing 3 qumodes (red lines) indexed by their frequency. A key point is that the carrier qumode at $\omega_o$ is coupled by PM to the input vacuum modes. When $\ket\psi$ is a coherent state, the output remains a product of coherent states.} 
\label{fig:classical}
\end{figure}

Phase modulation (PM) at frequency $\Omega$ of a monochromatic carrier field of frequency $\omega_o$ creates harmonic sidebands at frequencies $\om n=\om o+ n\Omega$ ($n\in\mathbb Z$) and of amplitudes the Bessel functions of the first kind
\begin{equation}
e^{i[\omega_ot+m \sin(\Omega t + \phi)]}
= e^{i\omega_ot}\sum_{n=-\infty}^{\infty}J_n(m)\,e^{in(\Omega t+\phi)},
\end{equation}
where $m$ is the modulation index and $\phi$ the PM phase. An effective quantum model of  PM~\cite{Capmany2010} is the multiport frequency-domain beamsplitter (\fig{classical})
Hamiltonian 
\begin{equation}\label{eq:H1}
H_\textit{PM} = \hbar \frac{\alpha}{\tau}  e^{- i \phi} \sum_{j=-\infty}^{\infty}  a_j  a_{j+\Omega}^\dagger + \text{H.c.},   
\end{equation}
where $\alpha=m/2$, $\tau$ is the interaction time in the phase modulator, and $a_j$ is the annihilation operator of the qumode of frequency $j$ in units of the qumode spacing, i.e., of the free spectral range (FSR) of the OPO. We take $\Omega$ in units of the FSR from \eq{H1} on.  Unitarity of quantum mechanics requiring there be as many input ports as there are output ports, PM of a single qumode will feature vacuum contamination at the sideband frequencies $\omega_\pm$, \fig{classical}. Such inputs are undesirable as random vacuum fluctuations decorrelate squeezed and entangled quantum states. 

\begin{figure}[t]
\includegraphics[width=\columnwidth]{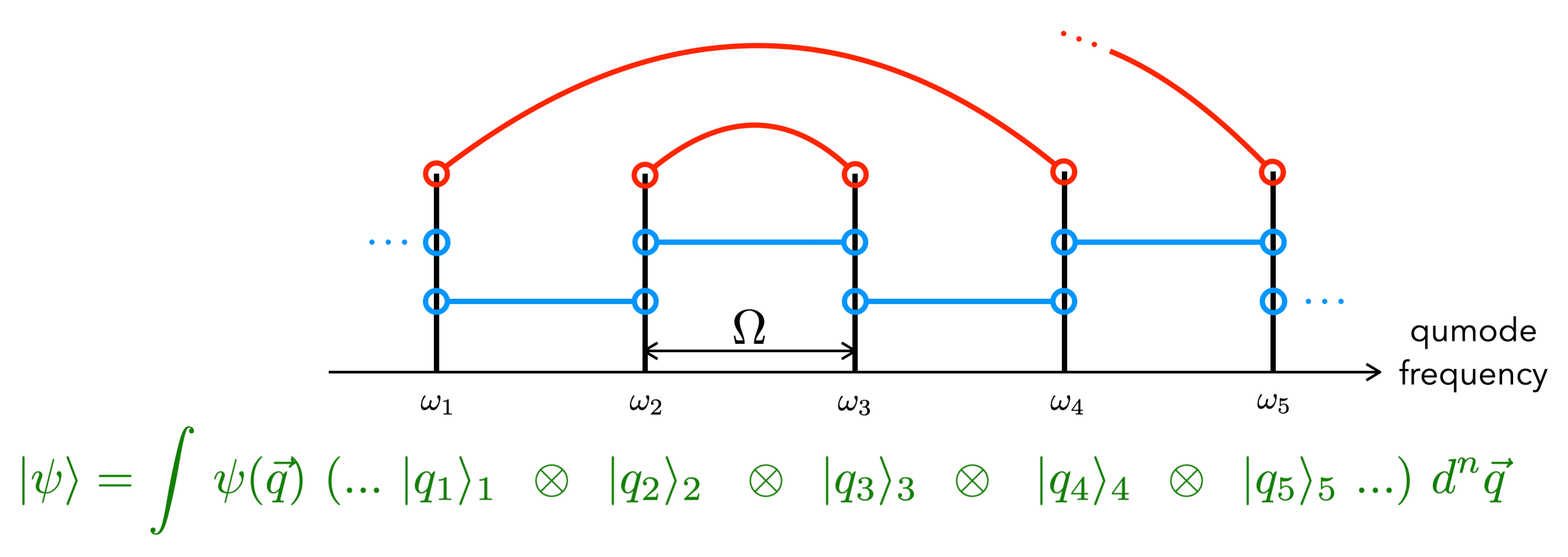}
\vglue-0.1in
\caption{Quantum description of PM at frequency $\Omega$ (blue lines) of the two-mode-squeezed (red lines, pump frequency $\om2+\om3$) QOFC: because the modulation frequency equals the qumode spacing, there is no vacuum input and a unitary operation can be realized solely on the initial comb state $\ket\psi$.}
\label{fig:quantum}
\vglue-0.2in
\end{figure}

The use of a QOFC input to PM replaces vacuum inputs with QOFC qumodes when $\Omega$ is an integer,  \fig{quantum}. Throughout this paper the input state will be a two-mode squeezed QOFC generated by a monochromatically pumped,  below-threshold OPO of Hamiltonian 
\begin{equation}\label{eq:tms}
H_\textit{TMS} = i\hbar\frac{r}{\tau'}\sum_{j=1}^{N}
a^{\dag}_{j}a^{\dag}_{p-j}+ \text{H.c.},
\end{equation}
where $p$ is the pump frequency in FSR units, $r$ is the squeezing parameter and $\tau'$ approximates the OPO cavity lifetime.\footnote{Note that a full treatment of the OPO spectrum can be done using the input-output formalism~\cite{Gardiner2004} but it is not required here to capture the essential physics.} This Hamiltonian generates two-mode squeezing~\cite{Ou1992} between qumodes $j$ and $p$-$j$, thereby creating an arbitrarily good approximation of an EPR pair~\cite{Einstein1935}.

In \sec{graphs}, we  derive the $N$-qumode quantum state created by two different quantum evolutions: first, by an externally modulated OPO, for which the output state is
\begin{align}\label{eq:casc}
\ket{\text{out}}=\exp(-\tfrac {i\tau}\hbar H_\textit{PM})\exp(-\tfrac{i\tau'}\hbar H_\textit{TMS})\ket0^{\otimes^N},
\end{align} 
second, by an intrinsically phase-modulated OPO (e.g.\ with  intracavity electro-optic modulation---EOM), described by 
\begin{align}\label{eq:int}
\ket{\text{out}}=\exp[-\tfrac {i\tau'}\hbar (H_\textit{TMS}+H_\textit{PM})]\ket0^{\otimes^N}.
\end{align}
One final word about the Hamiltonians of \eqs{H1}{tms}: we took the number of qumodes to be infinite in the former and finite in the latter, taking into account the phasematching bandwidth. This raises the question of qumodes at the edge of the phasematching bandwidth being coupled to vacuum modes which will degrade the entanglement. In addition, the edge of the phasematching bandwidth isn't sharp from one mode to its nearest neighbor so we can expect some variations of the entanglement in that region. While these concerns are legitimate, the resulting graph imperfections will remain confined to the boundary of the cluster state, due to its local nature, i.e. the ends of a 1D chain, the perimeter of a 2D graph, and the enveloping area of a 3D graph. Since, for cluster states, any local imperfections can be removed by single-qubit/qumode measurements~\cite{Briegel2001,Zhang2006,Gu2009}, these defects can be efficiently erased compared to the computation requirements which involve measurements of the bulk of the graph. We now outline the derivation procedure.

\subsection{Output state derivation}\label{sec:deriv1}

We proceed in three steps: first, deriving the symplectic matrix $\bf S$ of the quantum evolution in the Heisenberg picture, second, deducing from it the covariance matrix $\mathbf\Sigma$ of the state, third, obtaining the adjacency matrix $\bf V$ and error matrix $\bf U$ of the corresponding graph state.
An additional fourth step is warranted in the case of cascaded Hamiltonians, \eq{casc}, which employs a M\"obius transformation.

\subsubsection{Quantum evolution and covariance matrix} 

Any $N$-mode Gaussian Hamiltonian (i.e., quadratic in the field variables, we ignore displacements) yields a linear system of Heisenberg equations
\begin{align}
    \frac{d\mathbf x}{dt} = \mathbf G \mathbf x,\label{eq:HH}
\end{align}
where $\bf G$ is a symmetric matrix and where we posed {\bf x}=$({\bf Q}, {\bf P})^{T}$, {\bf Q}=$(Q_1,...,Q_N)^T$, {\bf P}=$(P_1,...,P_N)^T$. \Eq{HH}  can be solved by diagonalizing $\mathbf{G}=\mathbf{R}\mathbf{G}_\text{diag}\mathbf{R}^{-1}$, yielding the solution
\begin{align}
    {\bf x}(\tau) = \mathbf{S}\, {\bf x}(0),
\end{align}
where the symplectic matrix $\mathbf S$ is given by
\begin{align}
\mathbf{S} = \mathbf{R} \, e^{\tau\mathbf{G}_\text{diag}}\,\mathbf{R}^{-1}.
\end{align}
Note that, in a sequence of unitary operations, 
the symplectic matrix ordering is the Schr\"odinger picture one
\begin{equation}\label{eq:cascS}
    {\mathbf x}(\tau_n)={\mathbf S_n}{\mathbf S_{n-1}}\cdot\cdot\cdot {\mathbf S_1}{\mathbf x}(0)={\mathbf S}{\mathbf x}(0).
\end{equation}
Once $\mathbf{S}$ is known, we obtain the covariance matrix $\mathbf\Sigma$, which contains all information about the Gaussian state,
\begin{equation}\label{eq:CovFromSymp}
\mathbf{\Sigma} = \frac12\,\mathbf{S}\mathbf{S}^{T}.
\end{equation}
An important property of $\mathbf{\Sigma}$ is that it is related to the complex graph of the Gaussian state~\cite{Menicucci2011}, which we now briefly introduce. 

\paragraph{Reminders on graph and cluster states} 
A qubit graph state is canonically defined the following way~\cite{Briegel2001,Hein2004}: graph vertices $j$ denote qubits in state $\ket+_j=(\ket0_j+\ket1_j)/\sqrt2$ and graph edges $(j,k)$ denote controlled $Z$ gates $\text{CZ}_{jk}$. The stabilizers of a graph state---the operators that leave the state unchanged---form a group of Pauli generators $X_j\bigotimes_{k\in\mathbf N_j}Z_k$, where $j$=1,...,$N$ denotes each graph vertex and $\mathbf N_j$ is the neighborhood of $j$, i.e., the set of all vertices sharing an edge with $j$. A graph state can have any graph but the term ``cluster state'' is usually reserved for graph states that are sparse enough to allow measurement-based quantum computing~\cite{Bacon2009,Gross2009,Bremner2009}. In this paper, we use the ``cluster state'' moniker as the states we propose to generate are all sparse and adequate for one-way QC. 

The exact analogues of qubit graph states for CV quantum information are well defined~\cite{Zhang2006,Gu2009,Pfister2019}. A Gaussian cluster state is composed of vertices $j$ denoting phase-squeezed states
%\begin{align}
 $  S_j(r)\ket0_j
    = e^{\frac r2(a_j^{\dag2}-a_j^2)}\ket0_j,
$ %\end{align}
where $r$ is the squeezing parameter. Here ``phase'' pertains to the phase quadrature $P=(a-a^\dag)/(i\sqrt2)$. The edges $(j,k)$ of the cluster state graph denote the quantum nondemolition~\cite{Yurke1985} controlled-phase interaction $\exp(iQ_jQ_k)$, where $Q=(a+a^\dag)/\sqrt2$ is the field amplitude quadrature. In the limit of infinite squeezing, the CV graph state stabilizers are the generators of the Weyl-Heisenberg displacement group, e.g.\  $e^{-iP_j}\bigotimes_{k\in\mathbf N_j}e^{iQ_k}$. This leads to the equivalent definition of operators that multiply the state by 0, a.k.a.\ nullifiers: $P_j-\sum_{k\in\mathbf N_j}Q_k$. 
Finally, note that fault-tolerant QC is theoretically possible for finitely squeezed Gaussian states~\cite{Menicucci2014ft}. 

\subsubsection{Derivation of the Gaussian state graph}\label{sec:graph}

A qubit graph state is described by an unweighted adjacency matrix whose entries denote edge presence (value 1) or absence (value 0). Gaussian states are described by a complex-weighted~\footnote{Note that weighted graph states over qumodes are still stabilizer states, whereas they aren't over qubits.} adjacency matrix $\mathbf Z=\mathbf V +i\mathbf U$, where {\bf V} is the exact analog to the adjacency matrix of a qubit graph state~\cite{Pfister2019} and {\bf U} is the error matrix, which will be discussed in detail in \sec{error}. 

Any Gaussian state $\ket\psi$ can be defined by its nullifying operators, or nullifiers
\begin{align}\label{eq:Znullif}
    ({\bf P} - {\bf Z} {\bf Q})\ket\psi = \mathbf 0 \ket\psi,
\end{align}
which are, obviously, the logarithms of the stabilizers of $\ket\psi$. Because $\bf Z$ is complex in general, these  nullifiers are non-Hermitian and cannot be measured, which makes measurement-based quantum computing problematic since it must proceed by measuring, among other things, graph nullifiers~\cite{Gu2009}. Thus, only Gaussian states with $\bf Z$ real, i.e.\ with $\bf Z$=$\bf V$ and $\bf U$=$\mathbf0$, can be proper cluster states. 

However, measurement-based quantum computing is still possible with $\bf U$$\neq$$\mathbf 0$. In that case, we use the (arbitrarily good) approximate nullifiers given by Hermitian operators ${\bf P} - {\bf V} {\bf Q}$, which verify~\cite{Menicucci2011}
\begin{align}\label{eq:cov}
{\rm cov}[{\bf P} - {\bf V} {\bf Q}] & = \frac12\,{\bf U},
\end{align}
where we defined the covariance matrix in the standard way for vector operator  $\bf A$ with zero expectation values in state $\ket\psi$,
\begin{align}
    ({\rm cov}[\mathbf A])_{jk}=\frac12\bra\psi\{A_j^\dag,A_k\}\ket\psi.
\end{align}

We can deduce from \eq{cov} that an arbitrary Gaussian state of matrix $\bf Z$ can be accurately considered as a cluster state of matrix $\bf V$ {\em iff} 
\begin{description}
    \item{\em (i)}, the error matrix {\bf U} is diagonal~\cite{Gonzalez2020} and, \item{\em (ii)}, it verifies $\tr U \to 0$ in the limit of infinite squeezing~\cite{Menicucci2011}. 
\end{description}
In that case, $\mathbf P-\bf VQ$ are {\em squeezed}, uncorrelated Hermitian operators. Proper examination of $\bf U$ in light of requirements {\em(i,ii)} is therefore crucial and will be presented in \sec{error}.

We are now ready to determine the Gaussian graph matrix $\bf Z$ from the covariance matrix. The relation between the two is~\cite{Menicucci2011}
\begin{equation}\label{eq:CovMat}
\mathbf{\Sigma}_x ={\rm cov}[\mathbf x]= \frac12
	\begin{pmatrix}
	\mathbf{U}^{-1} & \mathbf{U}^{-1}\mathbf{V} \\
	\mathbf{V}\mathbf{U}^{-1} & \mathbf{U}+\mathbf{V}\mathbf{U}^{-1}\mathbf{V} 
	\end{pmatrix},
\end{equation}
where the block structure co\"incides with the definition of {\bf x}. One final remark highlights the importance of $\bf U$. The covariance matrix can be rewritten, using the rewritten symplectic vector $\mathbf y=(\mathbf Q,\mathbf P-\mathbf V\mathbf Q)^T$, as  
\begin{equation}\label{eq:CovMatY}
\mathbf{\Sigma}_y ={\rm cov}[\mathbf y]= \frac12
	\begin{pmatrix}
	\mathbf{U}^{-1} & \mathbb{O} \\
	\mathbb{O} & \mathbf{U}
	\end{pmatrix}.
\end{equation}
An allowable strategy for diagonalizing {\bf U} so that it verifies {\em (i,ii)} above is to apply {\em local} symplectic operations, which are equivalent to local unitaries (LU), to qumodes~\cite{Menicucci2011,Gonzalez2020} since these cannot change the state's entanglement. We will be making abundant use of this property. 

\subsubsection{Derivation of the graph for cascaded Hamiltonians [\eq{casc}]} \label{sec:deriv2}

As mentioned in \sec{QOPM}, we will consider PM either external or intrinsic to the OPO. Both options can be tackled using the procedure outlined above, with the former, cascading option making use of \eq{cascS}. Another, equivalent way to treat the cascading case is the M\"obius transformation~\cite{Menicucci2011}
\begin{equation}\label{eq:mobius}
{\bf Z'} = ({\bf C} + {\bf DZ})({\bf A} + {\bf BZ})^{-1}.
\end{equation}
where the second symplectic matrix is in the form
%\begin{equation}
${\bf S}=\left(\begin{smallmatrix} {\bf A} & {\bf B} \\ {\bf C} & {\bf D}\end{smallmatrix}\right)$.
%\end{equation}
Even though numerical calculations were used in this paper to solve the whole Heisenberg system, it is interesting to give, for reference, the analytic expression of the PM symplectic matrix 
\begin{equation}\label{eq:EOMsymp}
\mathbf S^{(PM)}=  \begin{pmatrix}\mathbf M & \mathbb O \\ \mathbb O & \mathbf M\end{pmatrix},
\end{equation}
where we have, for $\Omega=1$,
\begin{align}
\mathbf M_{jk} & = {\rm J}_{k-j}(m) - (-1)^{j}\, {\rm J}_{k+j}(m),
\label{eq:1}
\end{align}
for $\phi$=$\pi/2$ in \eq{H1}. This gives the well-known phase modulation spectrum, as was first obtained by Capmany and Fern\'andez-Pousa~\cite{Capmany2010}. Note that $\mathbf S^{(PM)}$ may not be block-diagonal for other values of $\phi$, which will lead to couplings between amplitude and phase quadratures. These are therefore totally controllable by setting $\phi$ experimentally.  

\section{Cluster state generation in the phase-modulated EPR QOFC}\label{sec:graphs}

\begin{figure}[ht!]  %[ht!]
\centerline{\includegraphics[width=.8\columnwidth]{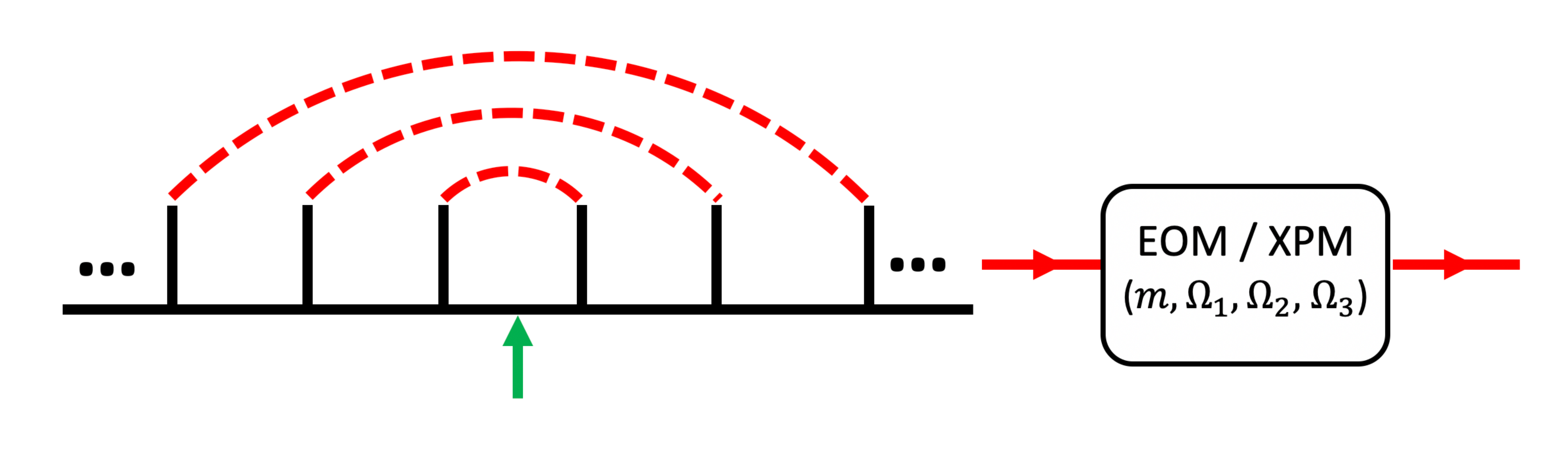}}
\caption{Phase modulation of a single QOFC. An OPO with a single pump frequency, whose half is denoted by the green arrow, creates TMS qumode pairs as indicated by the red dashed lines. Electro-optic phase modulation, or Kerr-medium cross phase modulation, is then done at index $m$ and frequencies $\Omega_{1,2,3}$.}
\vglue -.2in
\label{fig:config}
\end{figure}

We now demonstrate cluster state generation by external phase modulation of the EPR QOFC emitted by a monochromatically pumped OPO below threshold. We will also consider the case of the intrinsically modulated OPO at the end of the section. \Fig{config} depicts the experimental configuration.
A doubly resonant OPO is pumped at a single frequency $\omega_p$ such that frequency $\omega_p/2$ is set exactly halfway between 2 OPO mode frequencies (usually by a phaselock loop~\cite{Pysher2011,Chen2014}), as per the green arrow in \fig{config}. This generates independent EPR qumode pairs in two-mode-squeezed (TMS) states~\cite{Ou1992,Schori2002}, a.k.a.\ the EPR QOFC. While entanglement scalability is already present in this case, it manifests itself only as the scaling of the number of copies of a bipartite EPR state, rather than as the scaling of the size of a multipartite state. Phase modulation by the EOM of the OPO QOFC will change that: by modulating at one, two or three  frequencies, we can knit up 1D, 2D (square-grid), 3D (cubic) cluster states. We postulate that this extends to $n$-hypercubic cluster states, using $n$ modulation frequencies. 

Note that, for every graph presented in this paper, we conducted a thorough analysis of the errors due to finite squeezing and imperfections due to nonlocal modulation couplings. This analysis will be detailed in \sec{error}.

\subsection{Generation of 1D cluster states}\label{sec:J}

\begin{figure*}[ht!]
\centerline{\includegraphics[width=\textwidth]{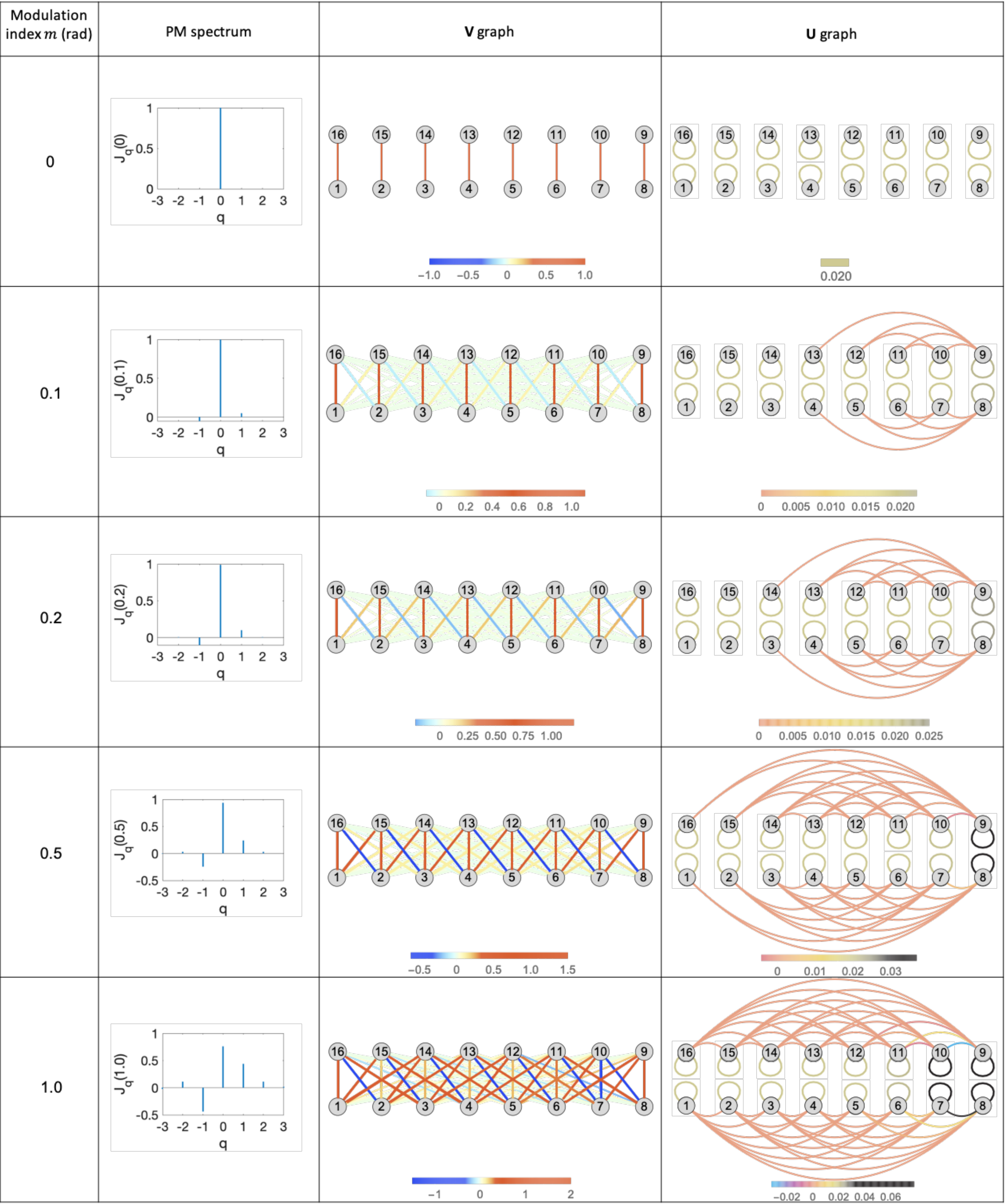}} 
\caption{$\mathbf{V}$ and $\mathbf{U}$ graphs for the state generated by OPO-extrinsic PM at $\Omega_1$=1 and $r$=2.3, for different $m$. The pump frequency is equal to the sum of the frequencies of all vertical qumode pairs in the $\bf V$ and $\bf U$ graph columns. Note that the mode labeled ``0'' in the PM spectrum column is any of the QOFC qumodes in the $\bf V$ and $\bf U$ graph columns. All self loops that have the same color in each $\bf U$ graph have a value of 0.02, regardless of the value of $m$.}
\label{fig:m_effect}
\end{figure*}

Following the steps in \secs{deriv1}{deriv2}, we derive the complex adjacency matrix $\bf Z$ of the created Gaussian state for $\Omega_1$=1, $r$=2.3, and for $m$=0, 0.1, 0.2, 0.5, and 1 rad. \Fig{m_effect} displays the real and imaginary parts of $\bf Z$, $\bf V$ and $\bf U$, after the appropriate LUs were applied; these LUs are Fourier transforms, i.e., rotations by $\pi/2$ in quadrature phase space $(Q,P)$, exerted on the lower-frequency-half qumodes, i.e., on all qumodes left of the green arrow in \fig{config}). 

We can immediately see that phase modulation yields multipartite entanglement, which is a first essential result of this paper.  
We now examine the particular graphs that can be generated by this method, turning first to the effect of the modulation index parameter $m$. The initial case $m=0$ corresponds to EPR pairs from the OPO with no phase modulation. When the latter is turned on, additional edges are created, whose weights increase with $m$, as the EPR weights decrease. The classical FM spectra in the left column of \fig{m_effect} give a good illustration of the effect on the quantum graph of the oscillations with $m$ of the Bessel-function amplitudes. 

\paragraph{OPO-intrinsic versus OPO-extrinsic PM}
In all cases presented in this paper, i.e., the 1D, 2D, and 3D graphs, we calculated both the extrinsic [\eq{casc}] and intrinsic [\eq{int}] cases. We found that both methods give the same graphs, but that the intrinsic method has a lower level of error, as described in \sec{error}. Because of the experimental simplicity of just placing an EOM after an OPO, we chose to present these extrinsic-OPO results in greater detail in \sec{graphs}, also thereby placing an upper bound on the imperfections.

Back to \fig{m_effect}, the onset of next-nearest neighbor couplings in the quantum graph co\"incides, unsurprisingly, with the growth of the first-order modulation sidebands, decrease of the carrier, and rise of the second harmonic sidebands. Such nonlocal graph edges are a known hindrance to one-way quantum computing~\cite{Bacon2009,Gross2009,Bremner2009}. However, we will show in \sec{error} that this problem can be circumvented upon closer, rigorous inspection of the edges' weights, which can always be found to be too small to be observable, when $m$ isn't too large. Note also that this optimal $m$ will decrease as the squeezing parameter $r$ increases. 

Remarkably, the imaginary part of the graph (rightmost column of \fig{m_effect}), given by the error matrix $\mathbf U$, complements the $\bf V$ graph, which clearly tends toward a complete bipartite graph with increasing $m$. In a sense, $\bf U$ leaves $\bf V$ globally invariant in terms of its bipartite structure, i.e., while $\bf V$ only connects qumode set $\{1,\dots,8\}$ to qumode set $\{9,\dots,16\}$ in \fig{m_effect}, $\bf U$ only connects qumodes within each of these two sets.  

The same property that spurious edges can always be made small enough will appear in the $\mathbf U$ graph. In \sec{error}, we will show that $\bf U$ is weak enough to validate this experimental technique. 

\begin{figure}[ht!]
\centerline{\includegraphics[width=0.75\columnwidth]{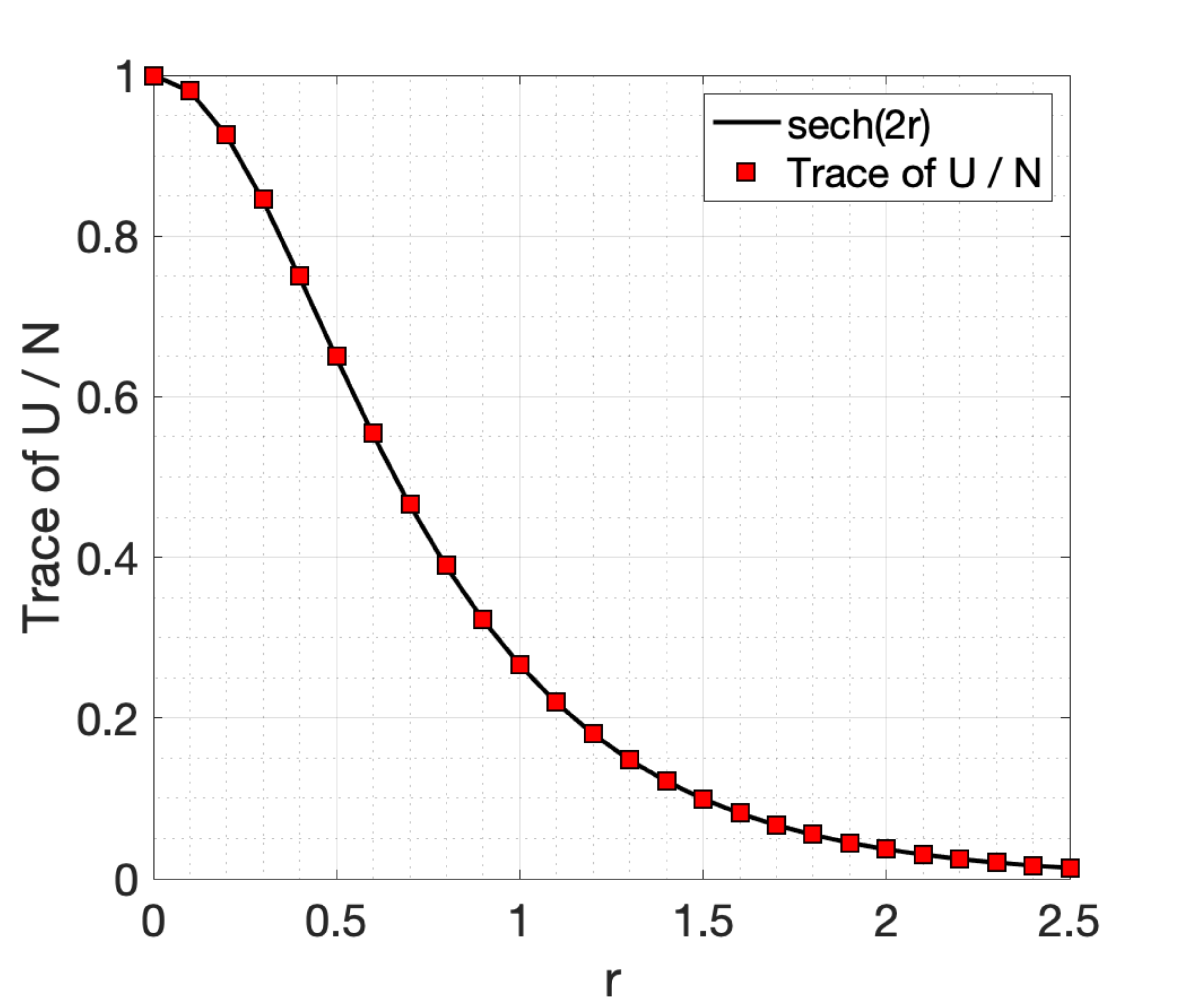}} 
%\vglue 1in
\caption{Plot of the average trace of $\bf U$ versus the squeezing parameter $r$, compared to the function $\text{sech}(2r)$. This $\mathbf{U}$ matrix is for the externally modulated QOFC at $\Omega_1$=1.}
\label{fig:TrU}
\end{figure}

Another crucial point that can be made immediately is that $\tr{\protect\mathbf U}\to0$ when $r\gg1$, as required and illustrated in \fig{TrU}, where $\tr{\protect\mathbf U}$ is plotted and shown to adhere to the theoretical value obtained analytically for two-mode squeezed states~\cite{Menicucci2011}. 
This point is important: it means that PM of two-mode-squeezed pairs doesn't detract from the fact that the overall cluster state error is solely determined by the amount of initial squeezing. One should indeed, remember that $\bf U$, being symmetric positive semidefinite, must tend to zero as a whole when its trace tends to zero as per requirement {\em(ii)}.\footnote{Note that requirement {\em(i)} remains crucial as we cannot reach this infinite squeezing limit experimentally and, again, a fault tolerance threshold has been proven to exist in this case of finite squeezing~\cite{Menicucci2014ft}.} 

\begin{figure}[ht!]
\centerline{\includegraphics[width=\columnwidth]{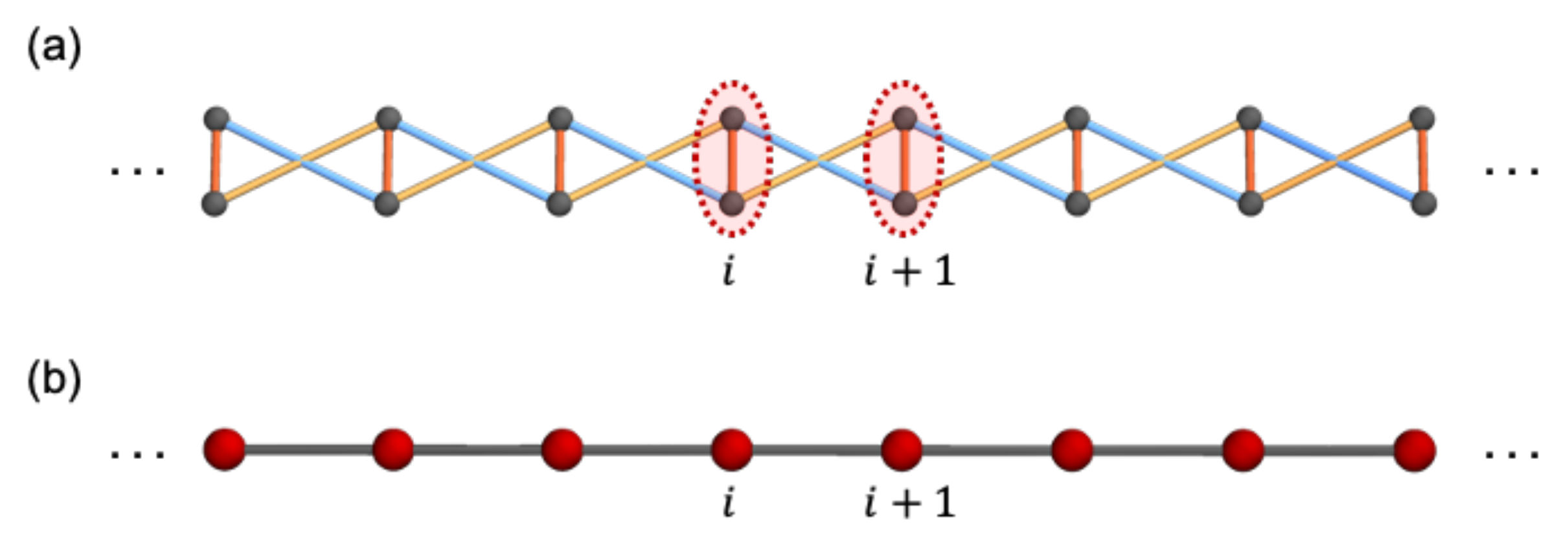}} 
\caption{(a), $\bf V$ graph of \fig{m_effect}, revealing its typical structure for 1D cluster state. Two modes that are in the red dashed circle are the EPR-qumodes. (b), compact representation of the graph using EPR macronodes.} 
\label{fig:ladder}
%\label{fig:TMSEOM2}
\end{figure}

At this point, we anticipate the conclusions of the rigorous error analysis of \sec{error} and ignore the weak edges in the whole complex graph to focus on the cluster states constructed by the dominant edges. We only have to limit the modulation index $m$ to low enough a value, keeping the PM couplings nearest-neighbor so as not to get nonlocal edges. The resulting graph is given by \fig{ladder}(a).
Swapping every other vertical pair of qumodes shows the graph to be a 1D ``ladder'' whose rungs are the initial EPR-pair qumodes. This dual graph structure connected by the initial EPR pairs will actually be a feature of the 2D and 3D graphs as well. In order to simplify the graph rendering in this case, we will bunch these EPR pairs into ``EPR macronodes,'' as in \fig{ladder}(b). This ladder state spans the whole phasematching bandwidth of the OPO, which can reach 10 THz in our previous work~\cite{Wang2014}. With a typical mode spacing of 1GHz~\cite{Pysher2011,Chen2014}, this yields on the order of $10^4$ entangled qumodes in this linear cluster state.

As is well known, the 1D graph isn't enough to generate the universal gate set in one-way quantum computing, for which a 2D graph is required.

\subsection{Generation of 2D cluster states}

The experimental configuration of \fig{config} is surprisingly versatile: just adding modulation frequencies allows to increase the dimension of the graph lattice. Again, we send the reader to \sec{error} for a detailed analysis of all graph imperfections and why they are negligible.

\begin{figure}[ht!]
\vglue 1in
\centerline{\includegraphics[width=
\columnwidth]{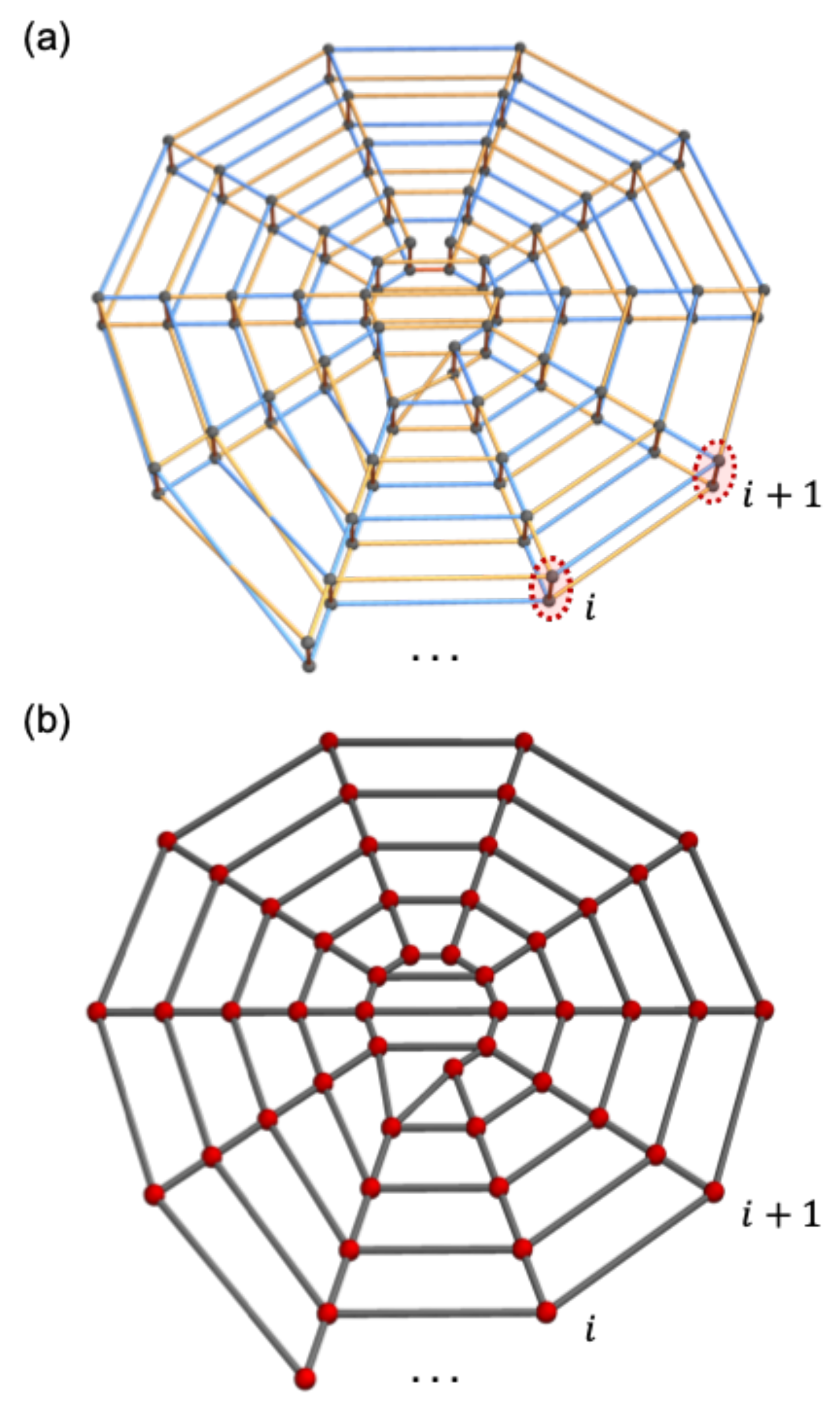}}
\caption{(a), square-grid cluster state  created with $\Omega_1$=1, $\Omega_2$=10.
Two modes circled by red dashed lines are EPR-qumodes, one of which is on the upper layer and the other one is on the bottom layer. (b), same graph, over EPR macronodes. The width of this square lattice is the number of ``spokes'' in the graph: $\Omega_2 / \Omega_1=10$.}
\label{fig:squaregrid}
\end{figure}

Driving the EOM with two modulation signals at frequencies $\Omega_{1,2}$ transforms \eq{H1} into
\begin{equation}\label{eq:H2}
H_{PM2} = \hbar \left[{\frac{\alpha_1}{\tau}  e^{- i \phi_1} \sum_{j=-\infty}^{\infty}  a_j  a_{j+\Omega_1}^\dagger +  \frac{\alpha_2}{\tau}  e^{- i \phi_2} \sum_{j=-\infty}^{\infty}  a_j  a_{j+\Omega_2}^\dagger } \right]+ \text{H.c.}   
\end{equation}
Here we set $\phi_1=\phi_2=\pi/2$, $\alpha_1=\alpha_2$, and follow the same procedure as in the 1D case, which yields the square-grid cluster state of \fig{squaregrid}. 
One can interpret this case in the following way: PM at frequency $\Omega_1$=1 creates next-neighbor coupling in the QOFC which forms a ladder graph; PM at frequency $\Omega_2$=10 then introduces additional coupling every 10 modes which is tantamount to spiraling the ladder into the two-layer square-grid cluster state of \fig{squaregrid}(a). As in \fig{ladder}(b), a more streamlined version of the graph can be obtained by considering EPR macronodes, \fig{squaregrid}(b).

This is an important result because the square-grid cluster state is a resource for universal quantum computing.\footnote{Note that this is true even though the cluster state is a Gaussian state and that universal quantum computing requires non-Gaussian resources for exponential speedup and quantum error correction. This is strictly equivalent to the qubit case where cluster states stabilized by Pauli operators, globally invariant under Clifford operations (just like CV clusters are stabilized by Weyl-Heisenberg displacements, globally invariant under Gaussian operations), even though non-Clifford resources are required to achieve exponential speedup. In both cases, the necessary respective non-Gaussian and non-Clifford gates are realized by like measurements on the cluster state.}

As was mentioned earlier, the width of the square lattice, which is the number of "spokes", is simply the ratio $\Omega_2/\Omega_1$, the total number $N$ of qumodes being determined by the phasematching bandwidth of the OPO's nonlinear medium. In the case of our previous experiments~\cite{Pysher2011,Chen2014}, for which we estimated $N\sim10^4$~\cite{Wang2014}, a 100 $\times$ 100 square cluster grid could therefore be created with PM at 1 and 100 GHz for a 1 GHz mode spacing. Note also that, in this case, the quasi-phasematching bandwidth can be further engineered to be larger.

\subsection{Generation of 3D cluster states}

At this point, we make the general claim that simply adding another modulation frequency  adds another dimension to the EPR-macronode graph, extending this procedure to yield hypercubic cluster graphs. We illustrate this in the 3D case, which is relevant to quantum computing because the 3D architecture is a known base for implementing topological error encoding over cluster states~\cite{Raussendorf2006}. 

\begin{figure}[ht!]
\centerline{\includegraphics[width=0.8\columnwidth]{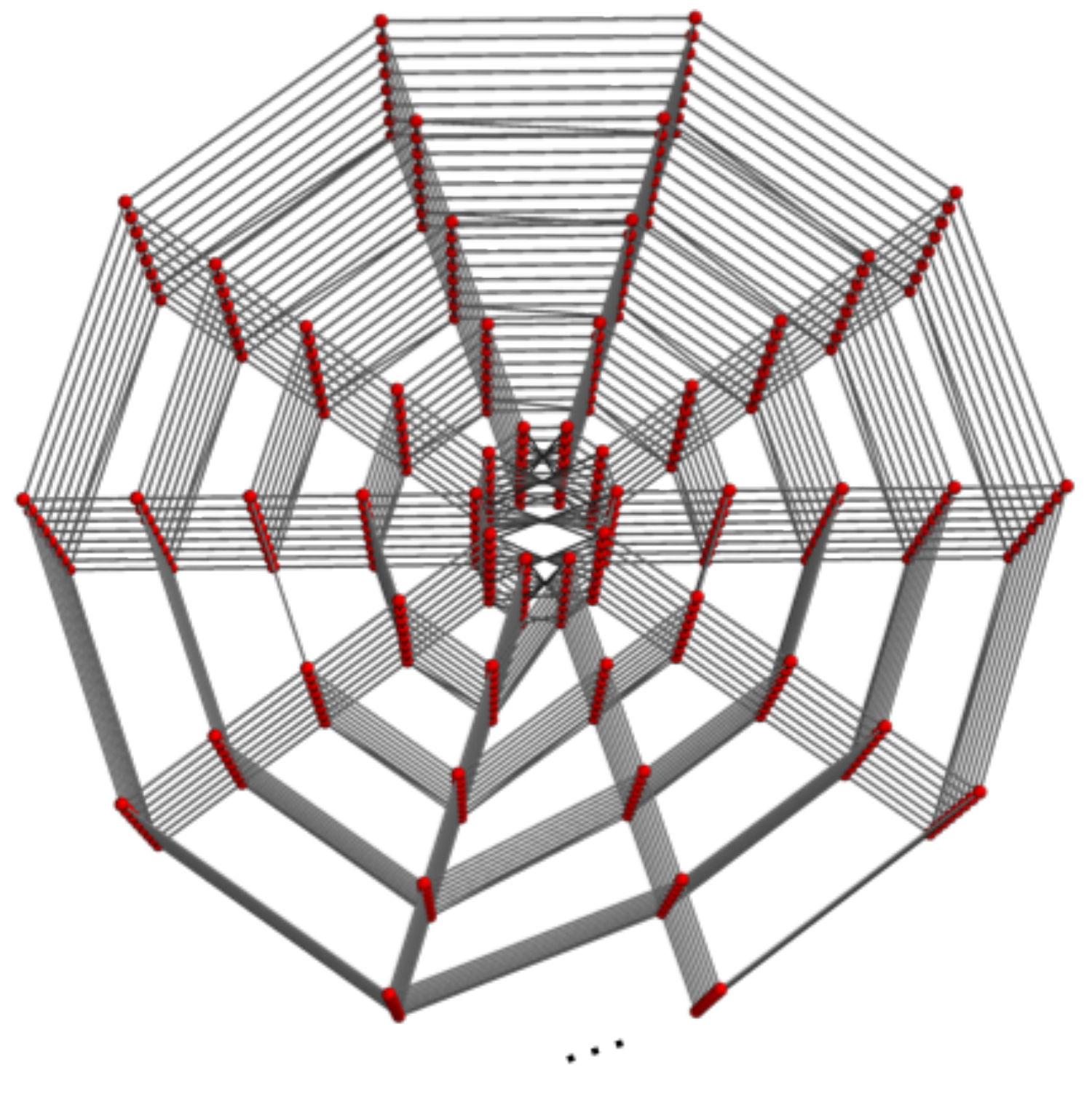}}
\caption{Cubic cluster state, over $N$=400 EPR macronodes, obtained from $\{\Omega_1,\Omega_2,\Omega_3\}=\{1,8,80\}$. Number of "spokes":10. Length: 8 macronodes. Spoke length: 5 macronodes.
}
\label{fig:cubic}
\end{figure}

With modulation frequencies $\{\Omega_1,\Omega_2,\Omega_3\}=\{1,8,80\}$, the quantum derivation yields the graph state pictured over EPR macronodes in \fig{cubic}.
The 6-edge valence of each graph vertex is clear in their vast majority. Note that, as always for cluster states, any local imperfections (graph center) in the graph can be removed by single-qumode measurements~\cite{Briegel2001,Zhang2006,Gu2009}. 

Finally, the ratio of PM frequencies determines the graph's, here cylindrical, structure: the number of ``spokes'' is set by $\Omega_3/\Omega_2$ and the length of the cylinder is set by $\Omega_2/\Omega_1$. The radius of the spokes increases with the mode number $N$ as $N/\Omega_3$. In the example of \fig{cubic}, a cubic cluster state is created over 400 macronodes in cylindrical shape with 10 "spokes", 5 set of macronodes in the radial direction, and a cylinder 8 macronodes long.

We now turn to the details of graph validation by scrutinizing the effects of the off-diagonal elements of $\bf U$.

\section{Graph error analysis}\label{sec:error}

As we mentioned earlier, an arbitrary Gaussian state is a valid cluster state iff 
\begin{description}
    \item{\em (i)}, the error matrix {\bf U} is diagonal~\cite{Gonzalez2020} and, \item{\em (ii)}, it verifies $\tr U \to 0$ in the limit of infinite squeezing~\cite{Menicucci2011}. 
\end{description}
While requirement {\em (ii)} has been systematically fulfilled in all previous realizations of CV cluster states~\cite{Yukawa2008a,Pysher2011,Chen2014,Yoshikawa2016,Asavanant2019,Larsen2019}, requirement {\em (i)} had not been considered until very recently~\cite{Gonzalez2020}, largely because all previous experimental realizations of cluster states had a diagonal $\bf U$. In this paper, $\bf U$ is not always diagonal, which can be seen as the price to pay for the considerable simplification of the experimental setup. We therefore  evaluate the contribution of off-diagonal elements of $\bf U$ and determine the precise conditions under which they can be neglected. 

Before we deal with error matrix $\bf U$, we first assume it fulfills both {\em (i)} and {\em (ii)} and focus  on the effect of the weak, undesirable edges in $\bf V$ that can be seen in \fig{m_effect} but not in \fig{ladder}.

\subsection{Effect of a spurious graph edge: bipartite case}

We first take the simplest example of a canonical cluster state: in the unphysical limit of infinite squeezing, a single qumode 1 has, by definition, nullifier $P_1$. In the realistic case of finite squeezing, the exact nullifier of qumode 1 in a single-mode squeezed (SMS) state of squeezing parameter $r_1$ is
\begin{align}
\mathcal N_{s1} &= S_1(r_1)\,a_1\,S_1^\dag(r_1) = P_1 -ie^{-2r_1}\,Q_1. 
\label{eq:Ns1}
\end{align}
These nullifiers are given by the complex $\bf Z$ graph, which always exists, but they are non-Hermitian, which makes the graph unsuited for quantum computing. A cluster state can be well defined as long as the imaginary part of $\bf Z$, $\bf U$, fulfills requirements {\em(i,ii)}. Finite squeezing is not a problem so long as it reaches a fault tolerant value, which has been theoretically proven to be within experimental reach ~\cite{Menicucci2014ft}. 

Two phase-squeezed qumodes coupled by gate ${\rm C_Z}=\exp(i\varepsilon Q_1Q_2)$ form a Gaussian cluster state of nullifiers
\begin{align}
\mathcal N_1 &= {\rm C_Z}\,\mathcal N_{s1}\,{\rm C_Z}^\dag = P_1 -ie^{-2r_1}\,Q_1 +\varepsilon\, Q_2\label{eq:N1}\\
\mathcal N_2 &= {\rm C_Z}\,\mathcal N_{s2}\,{\rm C_Z}^\dag = P_2 -ie^{-2r_2}\,Q_2 +\varepsilon\, Q_1, 
\label{eq:N2}
\end{align}
where $\bf U$ is diagonal and vanishes with increasing squeezing. We now ask the following question: if we wrongly assumed qumode 1 to be isolated when it is, in fact, linked to qumode 2 by a graph edge of small weight $\varepsilon$, how large could $\varepsilon$ be before its effects are detectable? 

To answer this question, we must first relate it to the actual physical measurements we can make on qumode 1. Under the assumption that we have two single-mode phase-squeezed states, the lowest measurement noise should be obtained by measuring the phase quadrature operator $P_1$, typically  by homodyne detection. In the case of a phase-squeezed qumode 1, observable $P_1$ has squeezed noise given by
\begin{align}
    (\Delta P_1)^2 = {}_1\!\bra0S_1(r_1)^\dag\, P_1^2\,S_1(r_1)\ket0_1 = \frac12\,e^{-2r_1}.\label{eq:dP}
\end{align}
We now assume $P_1$ when qumode 1 also has a $\rm C_Z$ graph edge of weight $\varepsilon$ with qumode 2 (squeezed by $r_2$), then we have 
\begin{align}
(\Delta P_1)^2 &= {\,}_{12}\!\bra{00}S_2^\dag S_1^\dag {\rm C_Z}^{\dag}\,P_1^2\,{\rm C_Z}S_1S_2\ket{00}_{12}\\
&= {\,}_{12}\!\bra{00}(P_1\,e^{-2r_1}-\varepsilon\, Q_2\,e^{2r_2})^2 \ket{00}_{12}\\
&= \frac12\,e^{-2r_1}\left[1+\varepsilon^2\, e^{2(r_1+r_2)}\right].\label{eq:dPe}
\end{align}
Comparing \eqs{dP}{dPe}, we deduce the condition for neglecting a graph edge of weight $\varepsilon$: 
\begin{align}
\varepsilon\ll \varepsilon_\text{min}=e^{-(r_1+r_2)} \label{eq:crit}
\end{align}
where $\varepsilon_\text{min}$ is the edge weight at which the quantum noise is raised by 3 dB on a single qumode quadrature measurement. 

We now connect this reasoning to the formalism of Gaussian graphical calculus~\cite{Menicucci2011} and, in particular, \eq{cov}. The procedure is the following: we define a ``trimmed'' version of the original graph {\bf Z}={\bf V}+i{\bf U} 
by zeroing all entries $\mathbf V_{jk}<\varepsilon_\text{min}$. In this case, it yields $\mathbf V'=\mathbb{O}$ and a diagonal error matrix
\begin{align}\label{eq:trim}
\mathbf U' 
&= 2\, \text{cov}[\mathbf P-\mathbf V'\mathbf Q]\\
&= 2\, \text{cov}[\mathbf P]\\
&= \begin{pmatrix} e^{-2r_1}\left[1+\varepsilon^2 e^{2(r_1+r_2)}\right] & 0\\ 0& e^{-2r_2}\left[1+\varepsilon^2 e^{2(r_1+r_2)}\right] \end{pmatrix}
\end{align}
with the general condition for $\tr {U'}\to0$ 
\begin{align}
\varepsilon\ll e^{-(r_1+r_2)},
\end{align}
which is identical to \eq{crit}.

\subsection{Effect of spurious graph edges: multipartite case}

As \eq{CovMatY} makes clear, an off-diagonal element of $\bf U$ has the general physical meaning of a correlation between two cluster-state nullifiers. Such covariances must be zero in order for the cluster state to be adequate for one-way quantum computing. However, we can derive a good quantitative estimate of the level at which such covariances can be neglected, for a given squeezing level ($\tr{\protect\mathbf U}$) of the graph. This estimate is the  error vector $\mathbf\Gamma$ defined by
\begin{align}
\Gamma_j = 
\frac1{\mathbf U_{jj}}\sum_{k\neq j}|\mathbf U_{jk}|.
\end{align}
Assuming $\mathbf U_{jj}$ is of the order of a squeezing factor $e^{-2r_j}$ (see \fig{TrU}), which ensures {\em (ii)}, then {\em(i)} can be fulfilled if each and every qumode $j$ satisfies
\begin{align}\label{eq:gamma}
\Gamma_j\ll1.    
\end{align}
This can even be relaxed a bit if one remember that local imperfections in cluster states can be measured out: if a majority of qumodes verify \eq{gamma}, then the few of them that don't can be taken out of the graph by measuring $Q_j$~\cite{Zhang2006,Gu2009}.

The procedure of \eq{trim} can be straightforwardly applied to the multipartite case: after zeroing out weak edges in $\bf V$, we seek to diagonalize $\bf U$ by applying LUs, here Fourier transforms, i.e., rotations by $\pi/2$ in phase space, and inspect the final $\mathbf U'$. If diagonalization wasn't successful (it's not always possible~\cite{Gonzalez2020}), we evaluate $\mathbf\Gamma$ to assess the closeness of the state to a perfect cluster state. 

Since we have noticed that $\bf U$ tends to be less and less diagonal as the dimensionality of the graph increases, we illustrate this procedure in the least favorable case of the cubic lattice produced by external PM. To obtain the graph of \fig{cubic} (as well as of \figs{ladder}{squaregrid}), all $\bf V$-edges weaker than a threshold value $\varepsilon_\text{min}$ were neglected. From this new $\mathbf V'$, a new error matrix $\mathbf U'$ was computed using \eq{trim}. 

\begin{figure}[ht!]
%\vglue 1in
\centerline{\includegraphics[width=0.8\columnwidth]{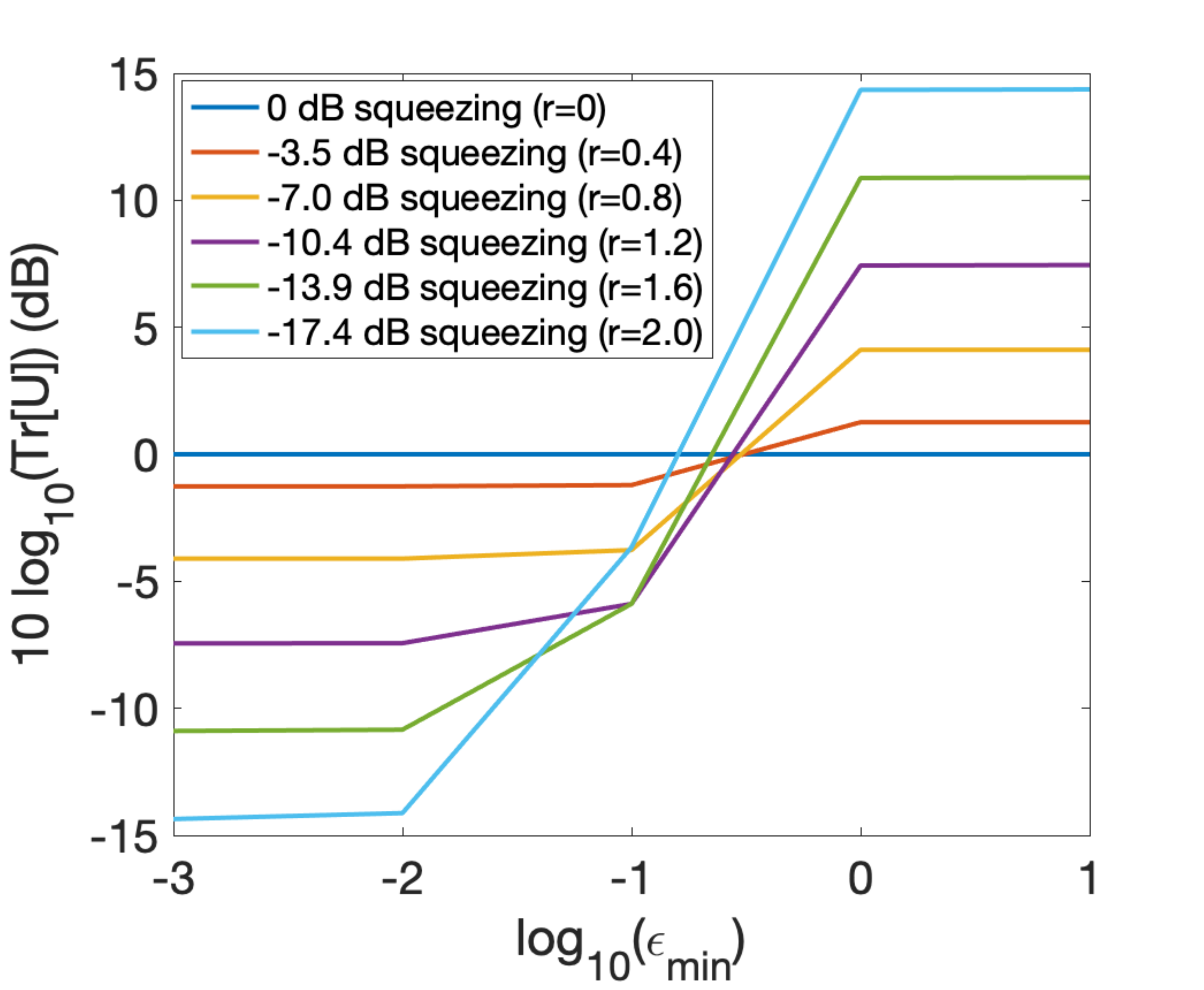}} 
\caption{Effect of graph ``trimming'' (removing all graph edges smaller than $\varepsilon_\text{min}$), on the trace of  $\mathbf{U}$, in the case of a cubic lattice produced by external FM. $\Omega_1=1$, $\Omega_2=8$, $\Omega_3=80$, $m=0.05$.} 
\label{fig:traceU}
\end{figure}

\Fig{traceU} displays the effect of such graph ``trimming'' on requirement {\em(i)}, i.e., $\tr{\protect\mathbf U}$ the cubic lattice cluster state, for different values of the squeezing parameter $r$.\footnote{Unsurprisingly, $\varepsilon_\text{min}$ must decrease when $r$ increases. This will require lowering $m$ in turn so as to keep the spurious edges weak.} 
The important conclusion from this graph is that there always exist an $\varepsilon_\text{min}$ such that requirement {\em(ii)} is fulfilled. Hence, graph trimming can always be performed, no matter how large the squeezing is, which ensures one can perform graph trimming above the fault tolerance threshold. 

\begin{figure}[ht!]
%\vglue 1in
\centerline{\includegraphics[width=0.8\columnwidth]{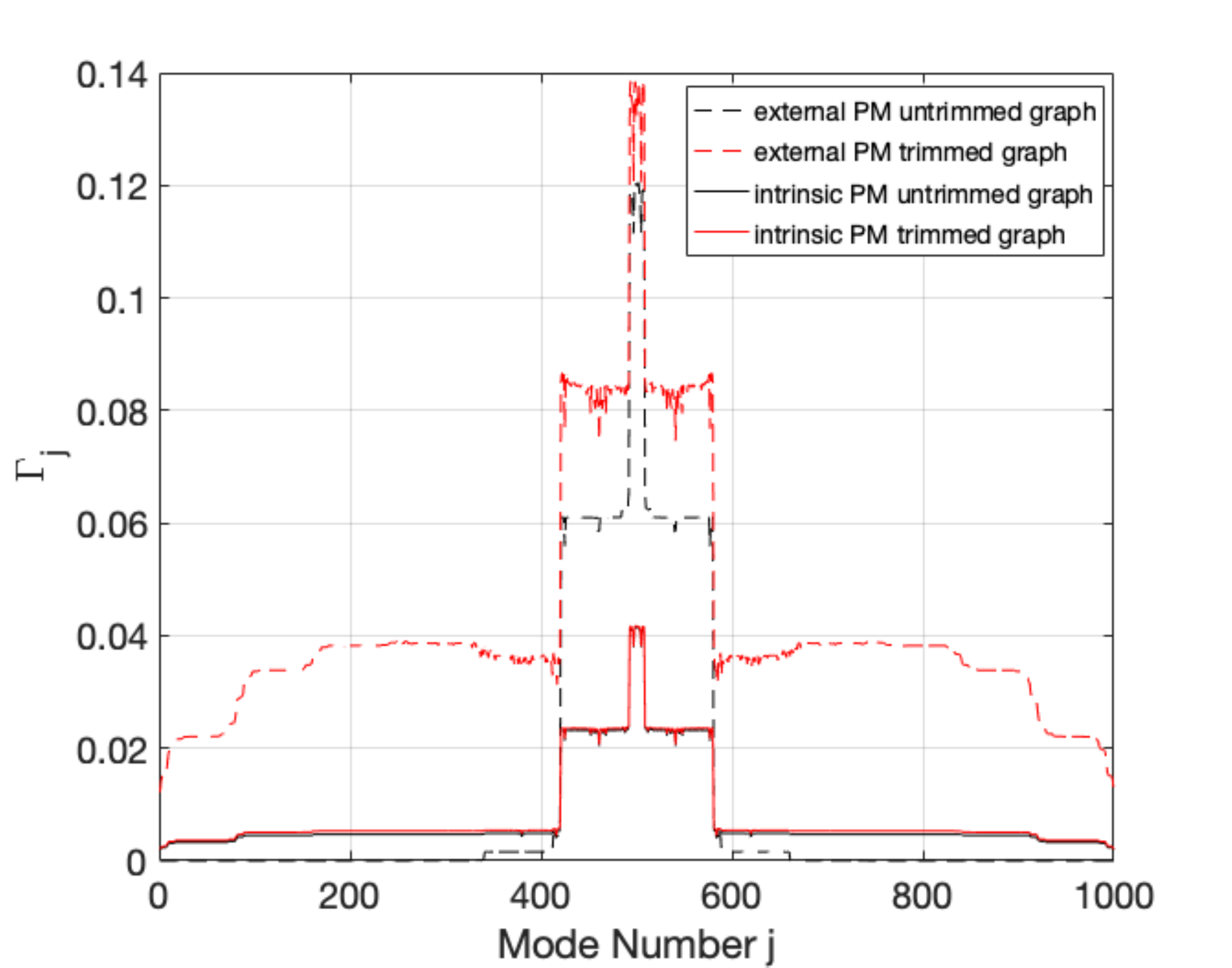}} 
\caption{$\mathbf\Gamma$ plot for the cubic lattice, for PM either external or internal to the OPO. $\Omega_1=1$, $\Omega_2=8$, $\Omega_3=80$, $m=0.05$, $r=1.2$, $\varepsilon_\text{min}$=0.01.} 
\label{fig:gamma}
\end{figure}

We now turn to the equally critical requirement {\em(i)} for the cubic graph state of \fig{cubic}. The components of error vector $\bf\Gamma$ of $\mathbf U'$ were computed for all 1000 qumodes in the state and plotted in \fig{gamma}, 
which addresses both cases of cascaded EPR pair generation and PM, and integrated PM in the OPO. In the former case (black dashed lines), only a small portion of the graph (the center of \fig{cubic}) has values of $\Gamma_j$ marginally larger than 0.1, and these can be measured out. Remarkably, the values of $\Gamma_j$ are close to zero for the vast majority of modes of the graph (0-400 and 600-1000), which corresponds to the bulk of the cubic lattice. Unsurprisingly, trimming the graph yields an increase of the error $\Gamma_j$ (red dashed lines) but, clearly, within manageable levels. 

In the case of PM intrinsic to the OPO (solid lines), the cubic graph is clearly ``cleaner'' from the start as the values of $\mathbf\Gamma$ are much lower and trimming has much less of an effect.

\section{Conclusion}

We demonstrated that the ``bare bone'' resources constituted by a monochromatically pumped, below-threshold  OPO along with phase modulation at multiples on the cavity spacing enable the generation of CV cluster states of arbitrary dimension, to arbitrarily low error level, compatible with the fault tolerance threshold predicted for CV quantum computing. The graph dimension is fully determined by the number and the ratio of phase modulation frequencies. For all squeezing levels, there exist modulation parameters that yield experimentally valid cluster states (see Supplemental Material). Of particular interest is the enhanced performance of phase modulation intrinsic to the OPO, as opposed to external to it. This experimental configuration is remarkably simple and compact and a marked simplification of all previous experimental realizations of large-scale cluster states, CV or otherwise. Note that these cluster states are deterministically and unconditionally generated, to the difference of other frequency-comb approaches that propose probabilistic linear-optics quantum computing~\cite{Lukens2017} or employ postselected photonic qubits~\cite{Kues2019}. 
Note that the frequency bandwidth of the fastest modulator involved dictates the final size of the generated cluster state. Even though we have used, throughout this paper, the example of an EOM for the phase modulator, even faster options exist such as  cross phase modulation in a Kerr medium, to which our analysis fully applies. Finally, the conceptual simplicity of this approach makes it well suited for implementations in integrated photonics. 

\section*{Funding}
This work was supported by NSF grants PHY-1820882, DMR-1839175, and EECS-1842641, and by BSF grant 2017743.

%\section*{Acknowledgments}

\section*{Disclosures}

The authors declare no conflicts of interest.

\section*{Supplemental Documents}
%\emph{Optica} authors may include supplemental documents with the primary manuscript. For details, see \href{http://www.opticsinfobase.org/submit/style/supplementary-materials-optica.cfm}{Supplementary Materials in Optica}. To reference the supplementary document, the statement ``See Supplement 1 for supporting content.'' should appear at the bottom of the manuscript (above the references).

\bigskip \noindent See \href{link}{Supplement 1} for supporting content.

%%%%%%%%%%%%%%%%%%%%%%% References %%%%%%%%%%%%%%%%%%%%%%%%%
%%%%%%%%%% If using BibTeX:
\bibliography{HypercubicComb}

\end{document}

\def\({\left(}
\def\){\right)}
\def\av#1{\left\langle #1\right\rangle}
\def\ad{a^\dagger}
\def\a2d{a^{\dagger 2}}
\def\b2d{b^{\dagger 2}}
\def\d#1{#1^\dagger}
\def\bydef{\stackrel{\wedge}=}
\def\eq#1{Eq.~(\ref{eq:#1})}
\def\Eq#1{Equation~(\ref{eq:#1})}
\def\eqs#1#2{Eqs.~(\ref{eq:#1}) \& (\ref{eq:#2})}
\def\eqlist#1#2{Eqs.~(\ref{eq:#1}-\ref{eq:#2})}
\def\Eqs#1#2{Equations~(\ref{eq:#1}) \& (\ref{eq:#2})}
\def\Eqlist#1#2{Equations~(\ref{eq:#1}-\ref{eq:#2})}
\def\sec#1{Section~\ref{sec:#1}}
\def\secs#1#2{Sections~\ref{sec:#1} \& \ref{sec:#2}}
\def\fig#1{Fig.\ref{fig:#1}}
\def\figs#1#2{Figs.\ref{fig:#1} \& \ref{fig:#2}}
\def\Fig#1{Figure~\ref{fig:#1}}
\def\Figs#1#2{Figures~\ref{fig:#1} \& \ref{fig:#2}}
\newcommand\tab[1]{Table~\ref{tab:#1}}
\newcommand\bra[1]{\left\langle\,#1\,\right|} %\mid doesn't work with \left, \right
\newcommand\ket[1]{\left|\,#1\,\right\rangle}
\newcommand\scalprod[2]{\left\langle\,#1\,\right|\left.#2\,\right\rangle}
\newcommand\om[1]{\omega_{\scriptscriptstyle #1}}
\newcommand\tr[1]{{\rm Tr}[\mathbf #1]}
\newcommand{\op}[1]{\textcolor{red}{[OP: #1]}}
 % Olivier's comments

%\graphicspath{{Figures/}}

\title{Supplemental Material for ``Hypercubic cluster states from the phase modulated quantum optical frequency comb''}

\author{Xuan Zhu}
\author{Chun-Hung Chang}
\author{Carlos Gonz\'alez-Arciniegas}
\affiliation{Department of Physics, University of Virginia, 382 McCormick Road, Charlottesville, VA 22903, USA}
\author{Avi Pe'er}
\affiliation{Department of Physics and BINA Center of Nano-technology, Bar-Ilan University, 52900 Ramat-Gan, Israel}
\author{Jacob Higgins}
\affiliation{Department of Physics, University of Virginia, 382 McCormick Road, Charlottesville, VA 22903, USA}

\author{Olivier Pfister}
\affiliation{Department of Physics, University of Virginia, 382 McCormick Road, Charlottesville, VA 22903, USA}
\email[olivier.pfister@gmail.com]{olivier.pfister@gmail.com}

\date{\today}

%\author{Xuan Zhu,\authormark{1,*} Chun-Hung Chang,\authormark{1}, Carlos Gonz\'alez-Arciniegas,\authormark{1} Avi Pe'er,\authormark{2} and Olivier Pfister\authormark{1,\dag}}

%\address{\authormark{1}Department of Physics, University of Virginia, 382 McCormick Road, Charlottesville, VA 22903, USA\\ \authormark{2}Department of Physics and BINA Center of Nano-technology, Bar-Ilan University, 52900 Ramat-Gan, Israel}

%\email{\authormark{*}xz7ab@virginia.edu,\authormark{\dag}olivier.pfister@gmail.com} %% email address is required



\begin{abstract}
In this document, we present the calculation results of the error vector $\Gamma$ of $\mathbf{U}$ matrices for the 1D chain, 2D square lattice and 3D cubic cluster states generated by both the externally and intrinsically phase-modulated optical parametric oscillator. These results show that at both low and high squeezing, we can always get valid cluster states.
\end{abstract}

\pacs{}

\maketitle

%%%%%%%%%%%%%%%%%%%%%%%%%%  body  %%%%%%%%%%%%%%%%%%%%%%%%%%

%\section*{Supplemental Documents}
%\emph{Optica} authors may include supplemental documents with the primary manuscript. For details, see \href{http://www.opticsinfobase.org/submit/style/supplementary-materials-optica.cfm}{Supplementary Materials in Optica}. To reference the supplementary document, the statement ``See Supplement 1 for supporting content.'' should appear at the bottom of the manuscript (above the references).

\section{Cubic lattice} 
\begin{figure}[htb]
%\vglue 1in
\centerline{\includegraphics[width=0.8\columnwidth]{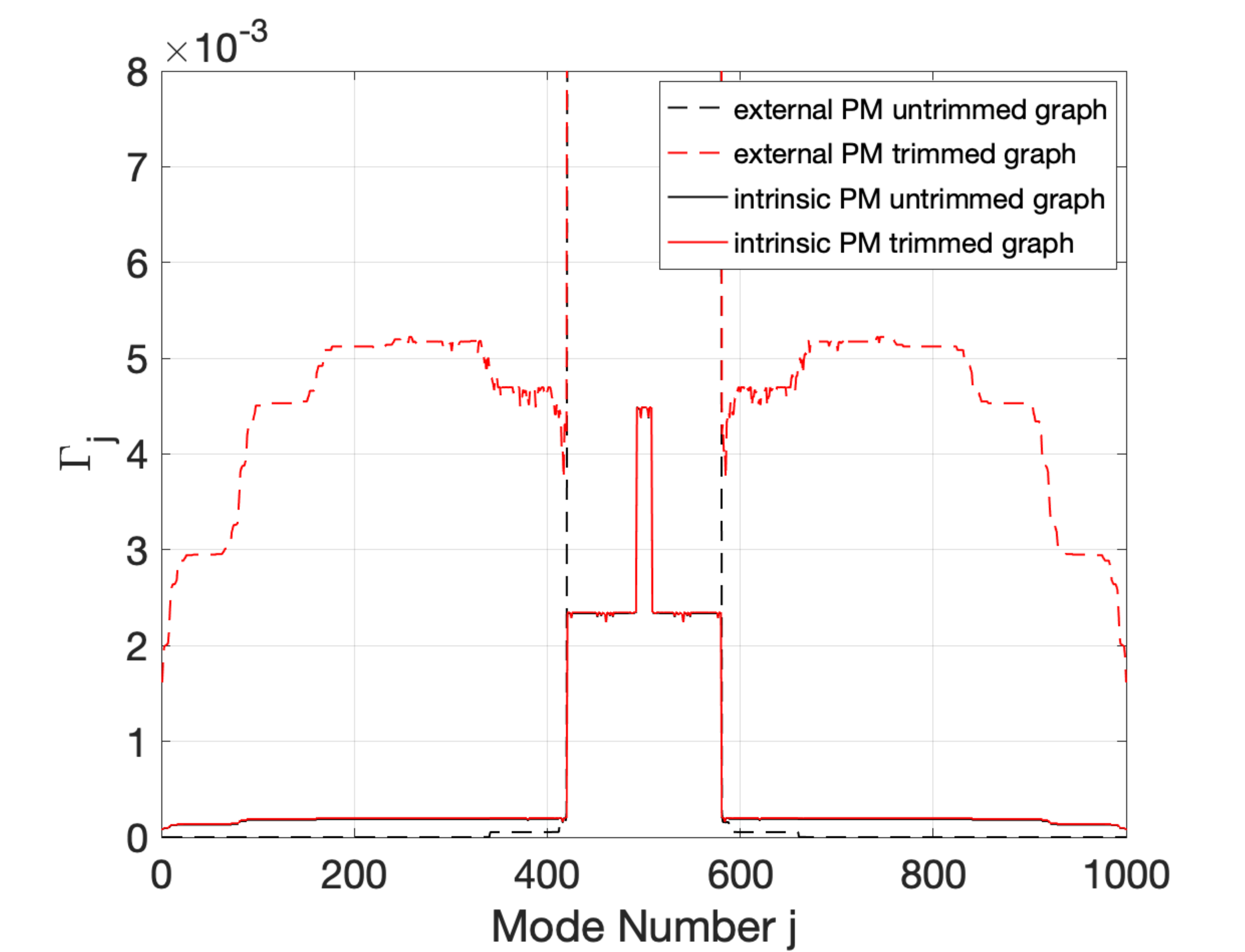}} 
\caption{$\mathbf\Gamma$ plot for the cubic lattice at $r=2.3$. $m=0.01$, $\varepsilon_\text{min}$ =0.001.} 
\label{fig:gamma3fmr23}
\end{figure}

\Figs{gamma3fmr23}{gamma3fmr04} are the $\bf\Gamma$ plots for cubic lattice generated with phase modulation frequencies $\Omega_1=1$, $\Omega_2=8$, $\Omega_3=80$. We can get the cubic lattice at $r$ equal to 2.3 and 0.4.

\begin{figure}[htb]
%\vglue 1in
\centerline{\includegraphics[width=0.8\columnwidth]{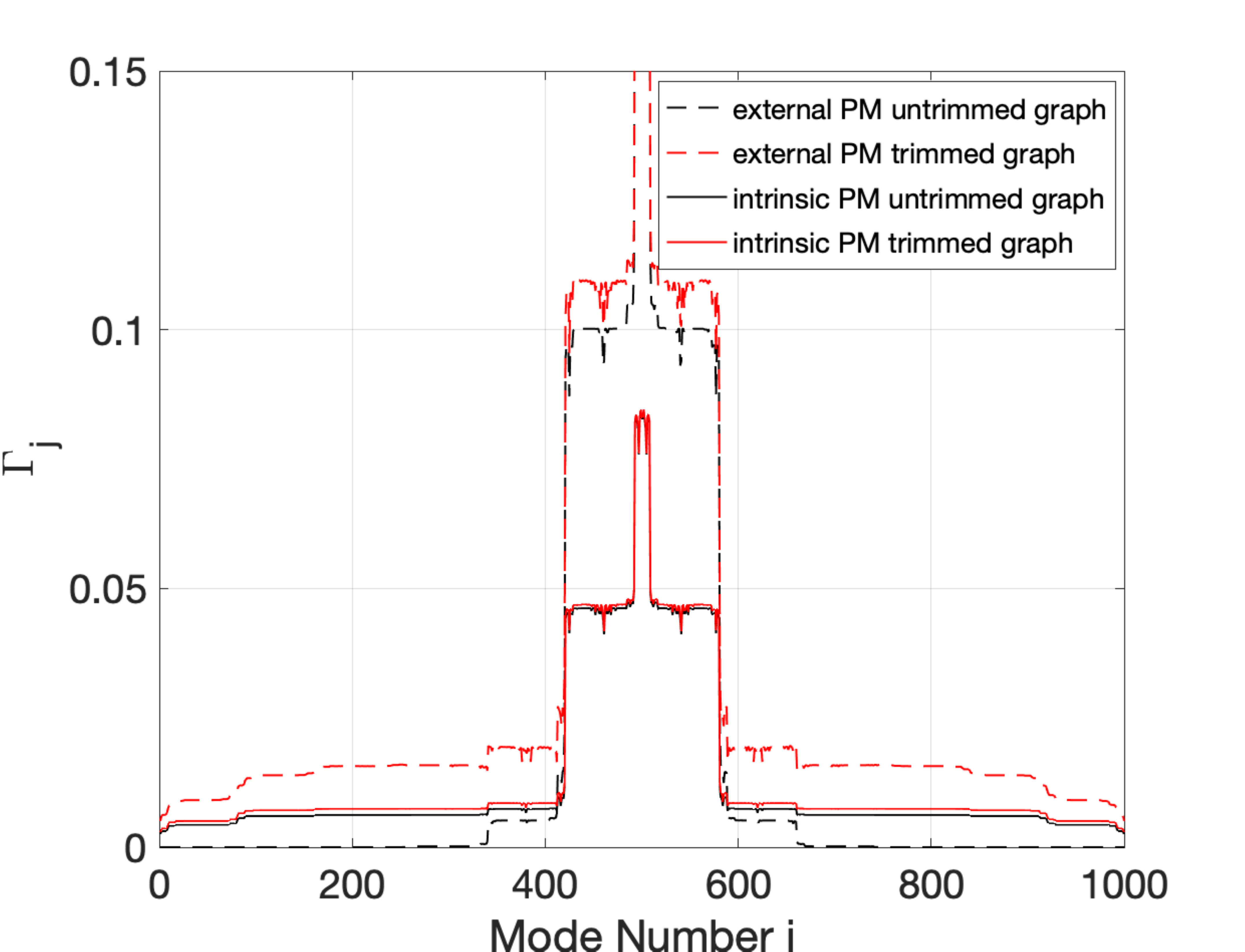}} 
\caption{$\mathbf\Gamma$ plot for the cubic lattice at $r=0.4$. $m=0.1$, $\varepsilon_\text{min}$ =0.01.} 
\label{fig:gamma3fmr04}
\end{figure}

\section{Square-grid lattice} 
\begin{figure}[H]
%\vglue 1in
\centerline{\includegraphics[width=0.8\columnwidth]{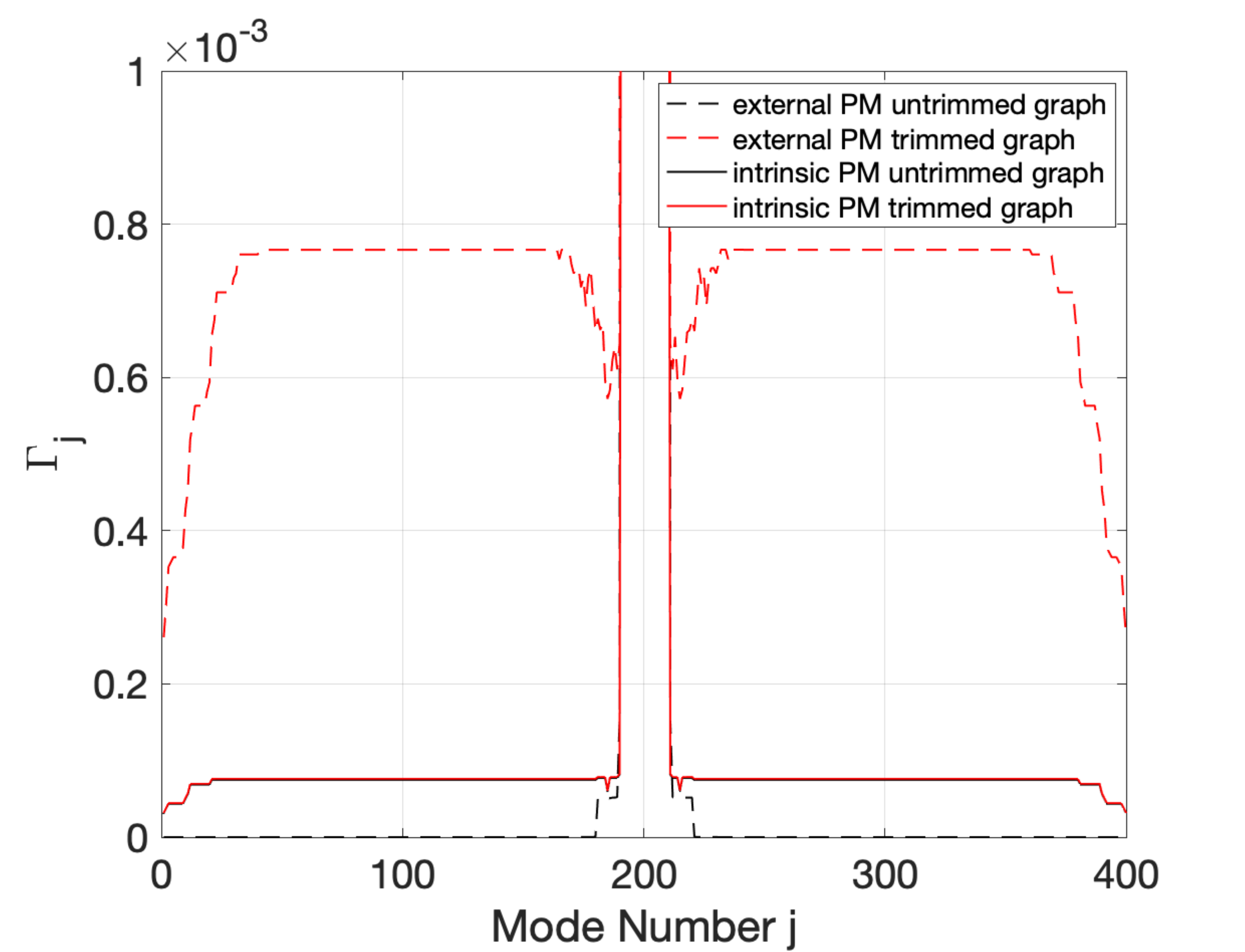}} 
\caption{$\mathbf\Gamma$ plot for the square-grid lattice at $r=2.3$. $m=0.01$, $\varepsilon_\text{min}$ =0.001.} 
\label{fig:gamma2fmr23}
\end{figure}

\begin{figure}[H]
%\vglue 1in
\centerline{\includegraphics[width=0.8\columnwidth]{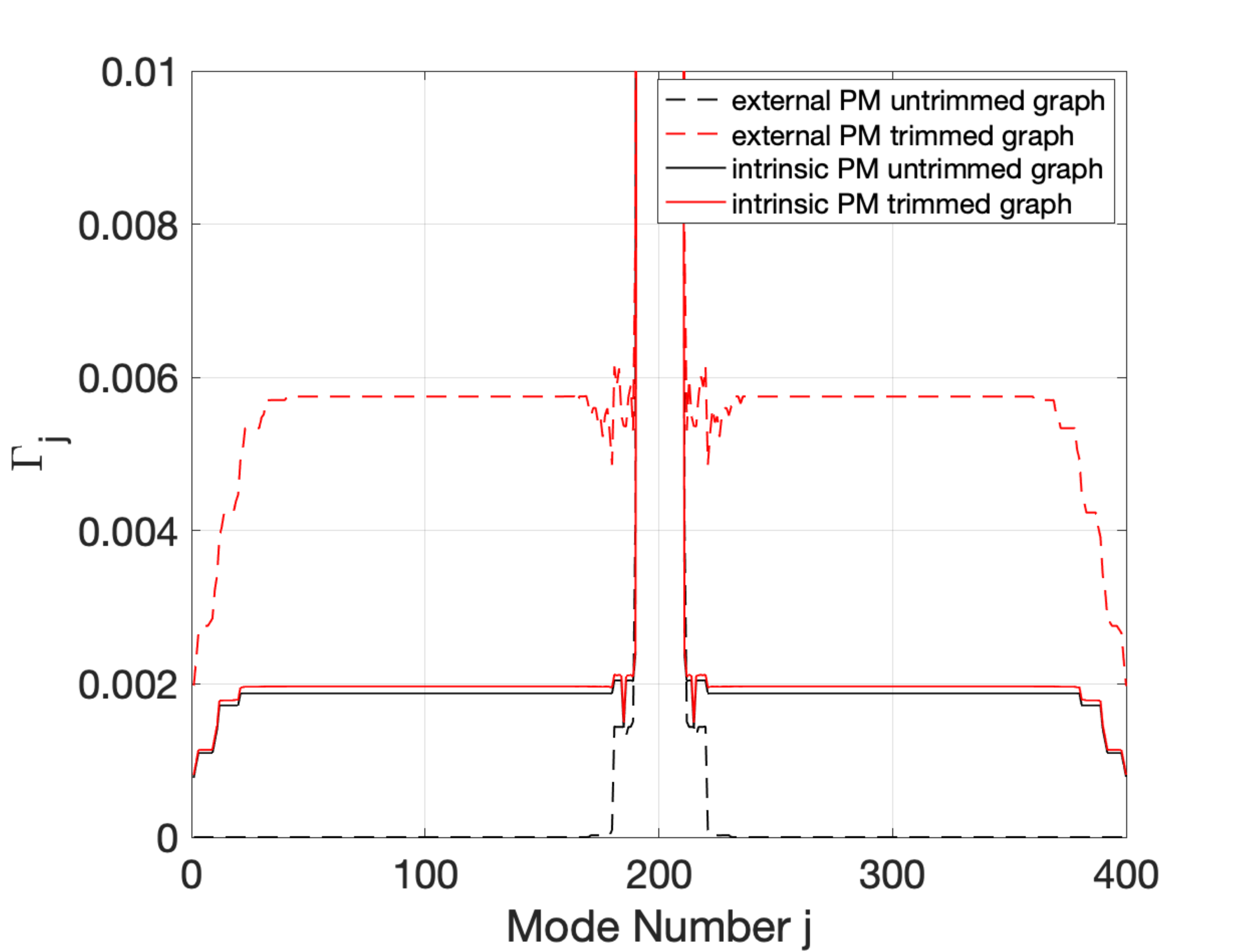}} 
\caption{$\mathbf\Gamma$ plot for the square-grid lattice at $r=1.2$. $m=0.05$, $\varepsilon_\text{min}$ =0.01} 
\label{fig:gamma2fmr12}
\end{figure}

\begin{figure}[H]
%\vglue 1in
\centerline{\includegraphics[width=0.8\columnwidth]{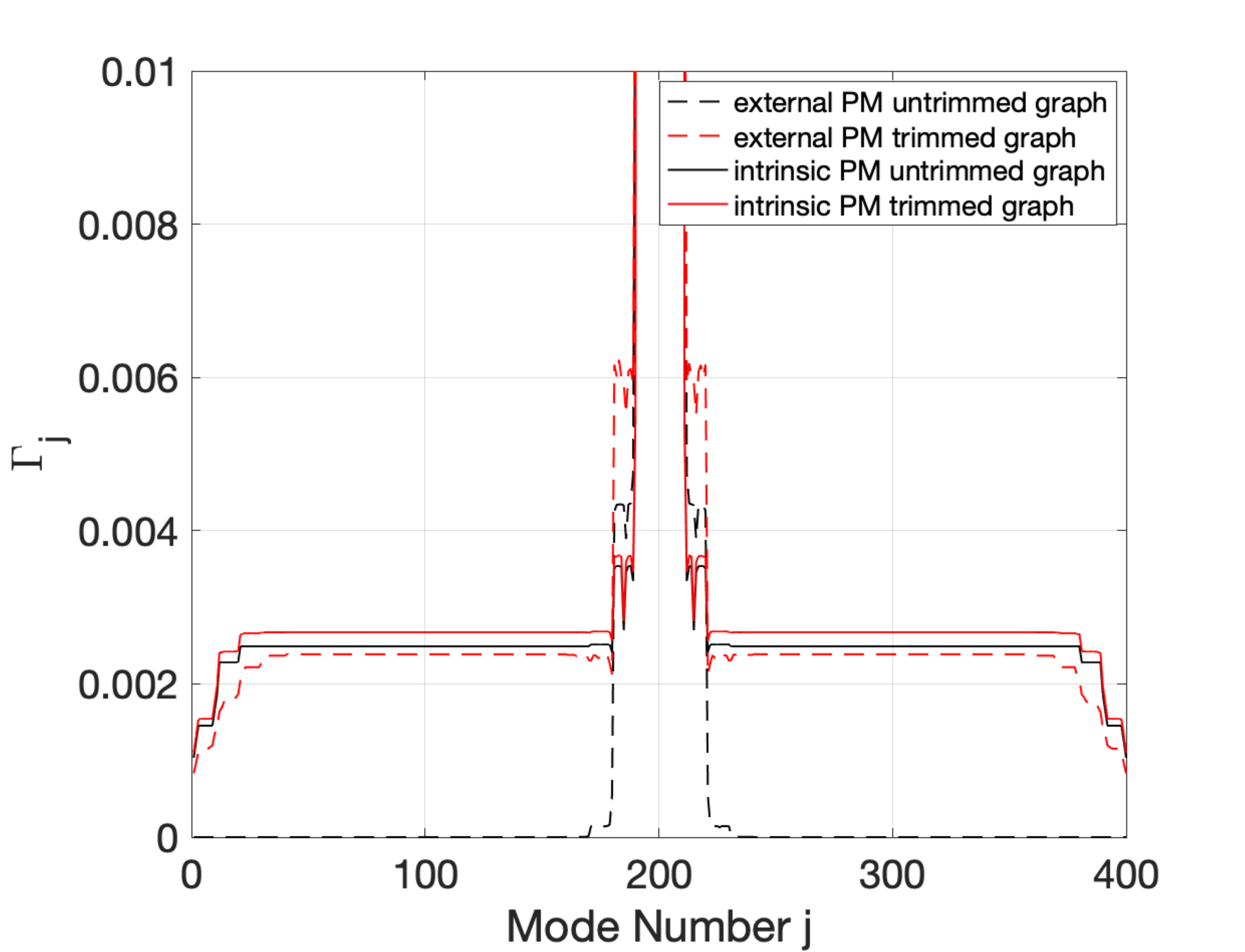}} 
\caption{$\mathbf\Gamma$ plot for the square-grid lattice at $r=0.4$. $m=0.1$, $\varepsilon_\text{min}$ =0.01} 
\label{fig:gamma2fmr04}
\end{figure}

\Fig{gamma2fmr23}, \fig{gamma2fmr12}, and \fig{gamma2fmr04} are the  $\bf\Gamma$ plots for square-grid lattice generated with phase modulation frequencies $\Omega_1=1$, $\Omega_2=10$. We can get the square lattice at $r$ equal to 2.3, 1.2 and 0.4.

\section{1D chain} 
\begin{figure}[h]
%\vglue 1in
\centerline{\includegraphics[width=0.8\columnwidth]{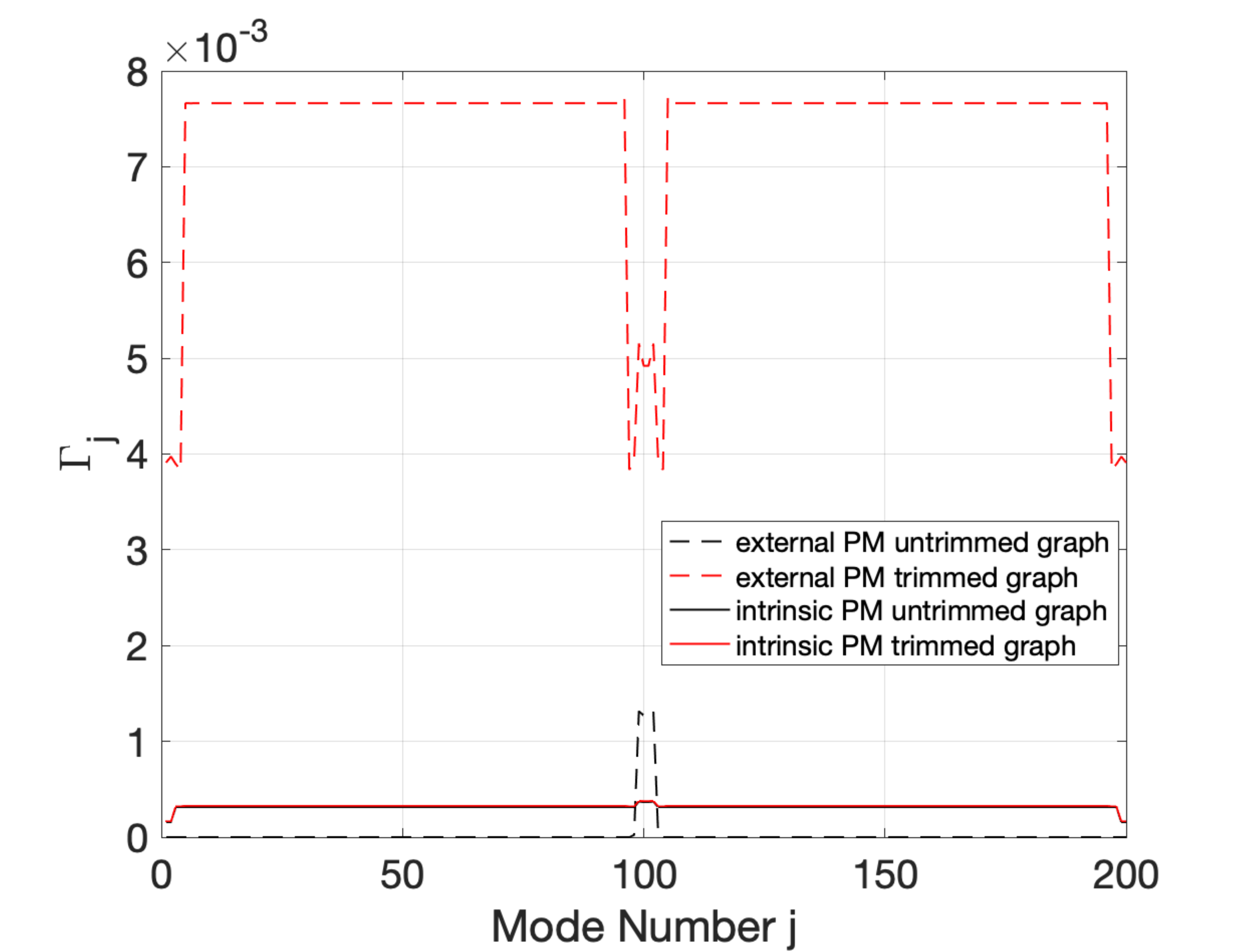}}
\caption{$\mathbf\Gamma$ plot for the chain cluster state at $r=2.3$. $m=0.05$, $\varepsilon_\text{min}$ =0.01} 
\label{fig:gamma1fmr23}
\end{figure}

\begin{figure}[H]
%\vglue 1in
\centerline{\includegraphics[width=0.8\columnwidth]{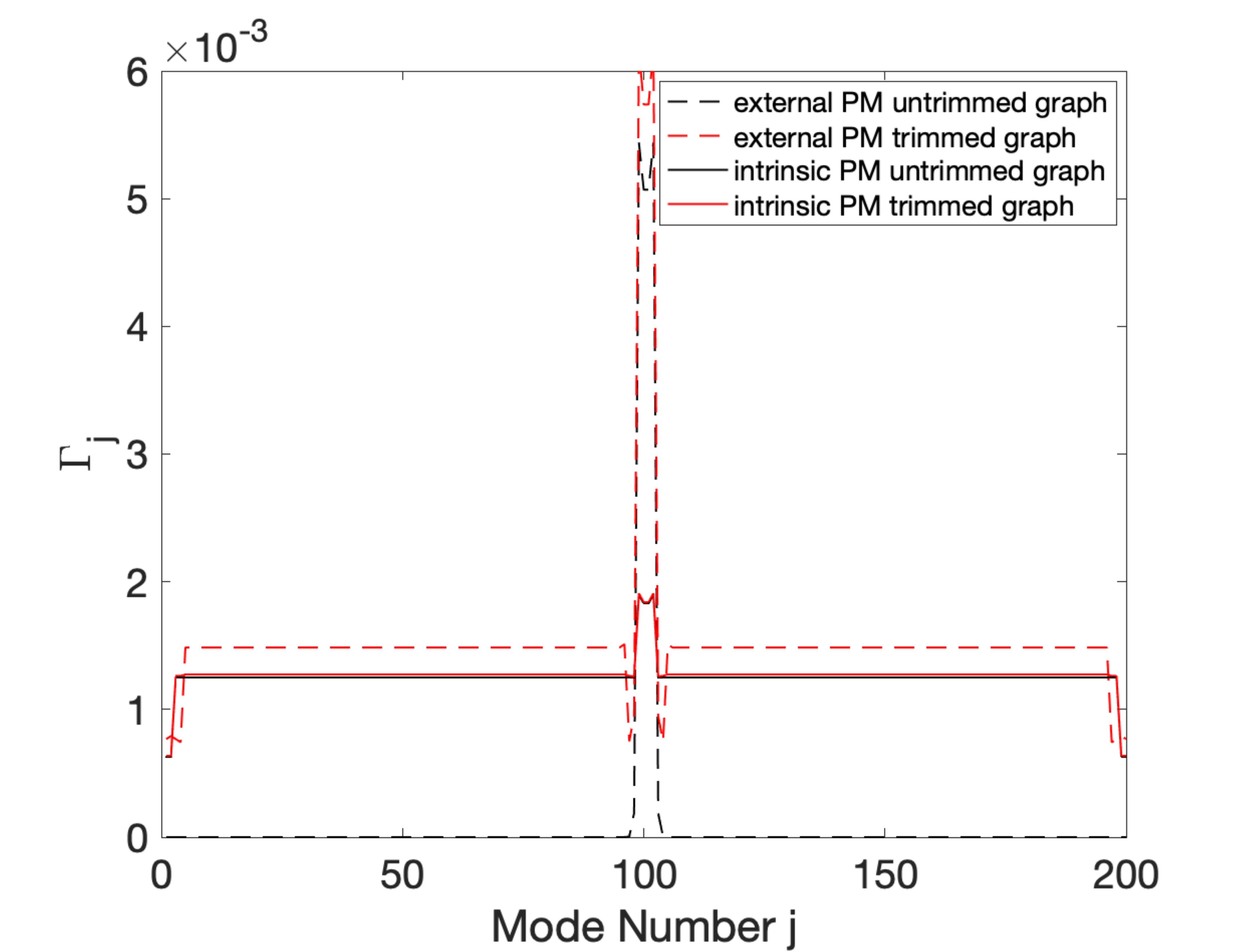}}
\caption{$\mathbf\Gamma$ plot for the chain cluster state at $r=1.2$. $m=0.1$, $\varepsilon_\text{min}$ =0.01} 
\label{fig:gamma1fmr12}
\end{figure}

\begin{figure}[H]
%\vglue 1in
\centerline{\includegraphics[width=0.8\columnwidth]{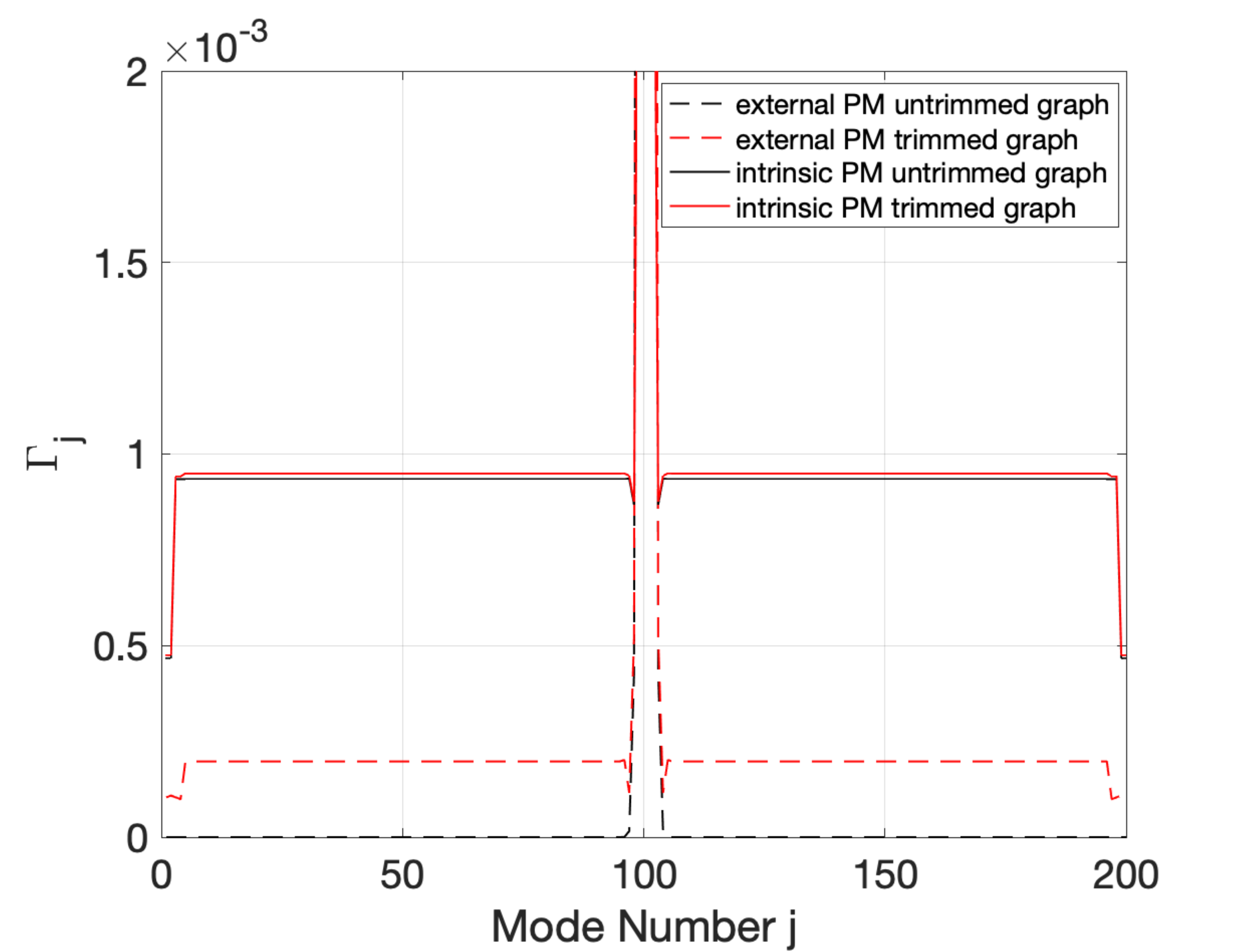}}
\caption{$\mathbf\Gamma$ plot for the chain cluster state at $r=0.4$. $m=0.15$, $\varepsilon_\text{min}$ =0.01} 
\label{fig:gamma1fmr04}
\end{figure}

\Fig{gamma1fmr23}, \fig{gamma1fmr12} and \fig{gamma1fmr04} are the  $\bf\Gamma$ plots for 1D chain cluster state generated with phase modulation frequencies $\Omega_1=1$. We can get the chain cluster state at $r$ equal to 2.3, 1.2 and 0.4.

%%%%%%%%%%%%%%%%%%%%%%% References %%%%%%%%%%%%%%%%%%%%%%%%%
%%%%%%%%%% If using BibTeX: